\documentclass[aps,showpacs,preprintnumbers,amsmath,amssymb,nofootinbib,eqsecnum,onecolumn,preprintnumbers]{revtex4}
\usepackage{graphicx}
\usepackage{epsf}
\usepackage{amsmath}
\usepackage{epstopdf}
\usepackage{bm}
\usepackage{color}
\usepackage{tabularx}
\usepackage{enumitem}
\usepackage{float}
\usepackage{array,booktabs}
\usepackage{footnote}
\usepackage{threeparttable}
\usepackage{graphicx}
\usepackage{hyperref}
\usepackage{amssymb,epsf}
\usepackage{latexsym}
\usepackage{epstopdf}
\usepackage{epsfig}
\usepackage{eurosym}
\usepackage{amsfonts}
\usepackage{amssymb}
\usepackage{xcolor}
\usepackage{subfigure}

\begin{document}

\title{Physical Properties of a Regular Rotating Black Hole:\\ Thermodynamics, Stability, Quasinormal Modes}
\author{S. H. Hendi$^{1,2}$\footnote{email address:
hendi@shirazu.ac.ir}, S. N. Sajadi$^{1}$\footnote{email address:
naseh.sajadi@gmail.com (Corresponding author)} and M.
Khademi$^{3}$\footnote{email address: Maryam.Khademi@obspm.fr}}
\affiliation{$^1$Physics Department and Biruni
Observatory, College of Sciences, Shiraz University, Shiraz 71454, Iran \\
$^2$Canadian Quantum Research Center 204-3002 32 Ave Vernon, BC V1T 2L7 Canada\\
$^3$Department of Physics, Shahid Beheshti University, G. C.,
Evin, Tehran 19839, Iran}

\begin{abstract}
Respecting the angular momentum conservation of torque-free
systems, it is natural to consider rotating solutions of massive
objects. Besides that, motivated by the realistic astrophysical
black holes that rotate, we use the Newman-Janis formalism to
construct a regular rotating black hole. We start with a
nonlinearly charged regular static black hole in the framework of
the standard general relativity and then obtain the associated
rotating solution through such a formalism. We investigate the
geometrical properties of the metric by studying the boundary of
ergosphere. We also analyze thermodynamic properties of the
solution in AdS spacetime and examine thermal stability and
possible phase transition. In addition, we perturb the black hole
by using of a real massless scalar field as a probe to investigate
its dynamic stability. We obtain an analytic expression for the
real and imaginary parts of the quasinormal frequencies. Finally,
we look for a connection between the quasinormal frequencies and
the properties of the photon sphere in the eikonal limit.
\end{abstract}
\maketitle

\section{Introduction}

One of the main open questions in theoretical physics is the existence of
singularity in different theories. From the gravitational point of view,
there might be something mysterious about the spacetime singularity. Thus,
investigation of such a pathology is a hot topic, either in physical and
mathematical communities or in philosophical circle.

The Einstein general relativity not only admits different solutions
including singularity, but expresses that such a singularity may be
unavoidable a real-world scenario. For this reason, one has to perceive the
nature of singularity to understand the nature of singular spacetimes.
Nonetheless, since the general relativity could not describe the nature and
physical properties of the spacetime singularity, one may look for
alternative viewpoint. The possibility of constructing a nonsingular
(regular) spacetime might be potentially important implication for avoiding
the breakdown of physical laws near the singularity, a region with extreme
curvature and vanishing volume. In addition, there is as yet no a consistent
theory of quantum gravity and some scientists believe that the singularity
would not occur in such a theory. Fortunately, the Einstein general
relativity allows some regular solutions without curvature singularity which
contain at most coordinate singularity \cite{Bardeen}. It is worth
mentioning that such regular black holes are not vacuum solutions of the
Einstein field equations. These regular solutions often include a special
class of nonlinear electrodynamics violating energy conditions in the
vicinity of the black hole \cite{Ayon-Beato1}-\cite{Sajadi:2017glu}. Till
date there has been a lot of significant work in regular solutions of
gravitating systems \cite{Bronnikov:2001tv}-\cite{Dymnikova:1992ux}.

On the other hand, according to the published results of the gravitational
wave observatories by the collaborations LIGO and VIRGO \cite%
{LIGO1,LIGO2,LIGO3} and shadow of black hole by Event Horizon Telescope \cite%
{Akiyama:2019cqa}-\cite{Akiyama:2019bqs}, one finds that the astrophysical
black holes are not static and spherically symmetric, but asymmetric due to
they have rotation. In other words, one of the most realistic features of
relativistic black holes is that they have angular momentum in a stationary
manner. Hence, in order to have a pragmatic black hole solution, one has to
consider rotating spacetime. However, introducing a new rotating black hole
solution, directly, is a nontrivial task as it turns out to be a rather long
process to obtain the Kerr solution. However, one may use the Newman-Janis
algorithm to convert static solutions to rotating ones \cite{Newman}-\cite%
{Modesto}. A special solution of the Kerr black hole in the presence of
electromagnetic field is obtained in \cite{Liao2017}. Phase transition of
the Kerr-Newman-AdS black hole with a model of dark energy is discussed in
\cite{Jafarzade2017} and its extension to nonlinear magnetic charge in \cite%
{Ndongmo2019}. The astrophysical aspects of rotating black holes, such as
shadow images and the geodetic precession frequency, have already been
studied in \cite{Haroon2019,Rizwan2019}. Besides that, other physical
properties of rotation black holes, in particular thermodynamic behavior and
photon sphere are of interest.

Taking into account the quantum effects near a black hole, one has to regard
it as a thermodynamical entity with a temperature and an entropy. Such a
statement help us to understand a deep connection among three interesting
theories; general relativity, quantum field theory and thermodynamics. In
other words, the black hole thermodynamics can be used as a bridge to
connect two apparently independent theories, general relativity and quantum
field theory. Black hole thermodynamics began, seriously, with the
pioneering works of Hawking and Beckenstein, and has recently become very
fascinating in the extended phase space by considering the cosmological
constant as a thermodynamic quantity. Taking a dynamical cosmological
constant $\Lambda=-8 \pi P$ into account, the consistent first law of black
hole thermodynamics and the associated Smarr relation are modified by
including a $PV$ term. Comparing such a modified first law of black hole
thermodynamics with that of everyday system, we find that in this
representation, the mass of the black hole $M$ is considered as the enthalpy
of the system instead of the internal energy \cite{kastor09}-\cite%
{kubiznak14}. Exploring the phase transition and critical behavior of the
black hole solutions in the extended phase space is another interesting
issue which is reported for different gravitating systems \cite%
{kubiznak12,Gunasekaran}-\cite{Hendi}. Thermal stability of a black hole
plays an important role in exploring its behavior near the equilibrium. It
is notable that a thermally stable black hole has a non-negative heat
capacity.

In addition to thermal stability criteria, one has to examine dynamical
stability of black holes under perturbations of the geometry and matter
fields. The robustness check of black holes against small perturbations is
sufficiently strong to veto some mathematical black holes. Regarding a
perturbative black hole, one may observe some oscillated behavior, named as
quasinormal modes (QNMs) which are related to some quasinormal frequencies
(QNFs). It is shown that QNMs are the intrinsic imprints of the black hole
response to external perturbations which means that such QNMs are
independent of initial perturbations. Authors of Refs. \cite%
{Nollert1993,Hod1998} show that the asymptotic behavior of QNMs is related
to the quantum nature of gravitation. It is also reported that for AdS black
holes, the imaginary parts of QNFs are corresponding to the perturbations
damping of a thermal state in the conformal field theory \cite%
{Horowitz2000,Cardoso2001}. So, the investigation of QNMs help us to find
the features of compact objects, the evolution of fields and also the
properties of spacetime \cite{Kokkotas1999,Berti2009,Konoplya2011}.

There are several approaches to the study black hole's QNMs. Ferrari and
Mashhoon \cite{vf}, working on the potential barrier in the effective
one-dimensional Schrodinger equation and obtain simple exact solutions. Such
a barrier is related to the photon sphere of the black hole \cite{yd}. We
should note that QNMs of regular black holes have been studied before \cite%
{sf}-\cite{scu}. The behavior of QNMs at the thermodynamics phase
transitions has been studied in \cite{Konoplya:2017zwo,Sajadi:2019hzo}.
Moreover, the relation between the QNFs and the thermodynamical quantities
at eikonal limit for static solution \cite{Hod:1998vk} and for rotating one
\cite{Musiri:2003ed} has been studied. Cardoso et al. \cite{Cardoso:2008bp}
showed that the real part of the QNMs is related to the angular velocity of
the last circular null geodesic while Stefanov et al. \cite{Stefanov:2010xz}
found a connection between black hole's QNMs in the eikonal limit and
lensing in the strong deflection limit. Furthermore, in Ref. \cite%
{Jusufi:2019ltj} the connection between the QNMs and the shadow radius for
static black hole and recently for rotating one \cite{Jusufi:2020dhz} has
been obtained. In this work, we use the Newman-Janis formalism to obtain a
regular rotating black hole solution. We also study the thermodynamic and
phase transition of such a rotating solution in the extended phase space by
using of standard approach. In addition, we study the stability of the black
hole against perturbation of spacetime and look for the QNMs \cite{cv,sc}.

The paper is organized as follows. In Sec. \ref{Sol}, we study the geometric
properties of an interesting class of regular rotating black hole. We study
the thermodynamics of rotating AdS black hole and look for possible phase
transition in Sec. \ref{Therm}. Section \ref{QNMs} is devoted to study the
QNMs and their connection to the properties of photon sphere in the eikonal
limit. The paper ends with our concluding remarks in Sec. \ref{Con}.

\section{Regular Rotating Black Hole\label{Sol}}

The $4-$dimensional action governing nonlinearly charged black holes in the
presence of a negative cosmological constant is given by%
\begin{equation}
\mathcal{A}=\dfrac{1}{16\pi }\int d^{4}x\sqrt{-g}\left[ \mathcal{R}+\dfrac{6%
}{l^{2}}-\mathcal{L(F)}\right]  \label{eqaction}
\end{equation}%
in which $g$ is the determinant of the metric tensor, $l=\sqrt{-3/\Lambda }$
denotes the AdS length related to the negative cosmological constant, $%
\mathcal{R}$ is the Ricci scalar and $\mathcal{L(F)}$ is an arbitrary
function of the Maxwell invariant $\mathcal{F}=\mathcal{F_{\mu \nu }}%
\mathcal{F^{\mu \nu }}$. Applying the variational principle to the action (%
\ref{eqaction}), one can show that the field equations are given by
\begin{equation}
G_{\mu \nu }-\frac{3}{l^{2}}g_{\mu \nu }=2\mathcal{L_{F}F}_{\mu \lambda }%
\mathcal{F}_{\nu }^{\lambda }-\dfrac{1}{2}g_{\mu \nu }\mathcal{L(F)}
\label{Einstein equation}
\end{equation}%
\begin{equation}
\nabla _{\mu }\left( \mathcal{L_{F}F}^{\mu \nu }\right) =0,
\label{Maxwell equation}
\end{equation}%
where in the above equations $G_{\mu \nu }$ is the Einstein tensor and $%
\mathcal{L_{F}}=d\mathcal{L}/d\mathcal{F}$.\newline
The metric of rotating charged regular black hole in the Boyer-Lindquist
coordinates is obtained as \cite{Newman}, \cite{Dymnikova}
\begin{equation}
dS^{2}=-\dfrac{\Delta _{r}}{\Sigma }\left( dt-\dfrac{a\sin ^{2}(\theta )}{%
\Xi }d\phi \right) ^{2}+\dfrac{\Sigma }{\Delta _{r}}dr^{2}+\dfrac{\Sigma }{%
\Delta _{\theta }}d\theta ^{2}+\dfrac{\Delta _{\theta }\sin ^{2}(\theta )}{%
\Sigma }\left( adt-\dfrac{r^{2}+a^{2}}{\Xi }d\phi \right) ^{2}
\label{metric}
\end{equation}%
where
\begin{eqnarray*}
\Delta _{r} &=&(r^{2}+a^{2})(1+\dfrac{r^{2}}{l^{2}})-2f, \\
\Sigma &=&r^{2}+a^{2}\cos ^{2}(\theta ), \\
\Xi &=&1-\dfrac{a^{2}}{l^{2}}, \\
\Delta _{\theta } &=&1-\dfrac{a^{2}}{l^{2}}\cos ^{2}(\theta )
\end{eqnarray*}%
and the functional form of $f(r)$ depends on the choice of the EM Lagrangian
$\mathcal{L(F)}$. Our approximate functional form of $f(r)$ is introduced in
\cite{Balart:2014cga}, \cite{Ghosh:2014pba} as%
\begin{equation}
f(r)=Mr\exp \left( -\dfrac{q^{2}}{2Mr}\right) .  \label{f(r)}
\end{equation}%
Here, for the sake of simplicity we have considered above function for $f(r)$%
. It is worth mentioning that the asymptotic behavior of the metric (\ref%
{metric}) is in agreement to the Kerr-Newman-AdS black hole.\newline
The radius of the horizon $r_{+}$ can be obtained from the following
equation
\begin{equation}
\Delta _{r}|_{r=r_{+}}=(r_{+}^{2}+a^{2})\left( 1+\dfrac{r_{+}^{2}}{l^{2}}%
\right) -2Mr_{+}\exp \left( -\dfrac{q^{2}}{2Mr_{+}}\right) =0.
\label{horizon}
\end{equation}%
The existence/nonexistence of real positive root of the above equation
indicates two scenarios, regular black hole or no-horizon solution. The
regular black hole and no-horizon cases may be separated by introducing the
extremal horizon of the black hole solution. The extremality condition is
defined by the following relation
\begin{equation*}
\Delta _{r}|_{r=r_{+}}=\Delta _{r}^{^{\prime }}|_{r=r_{+}}=0.
\end{equation*}%
Regarding $\Delta _{r}(r=r_{+})=0$, one can obtain%
\begin{equation}
M=\dfrac{q^{2}}{2r_{+}\mathcal{W}\left( \dfrac{l^{2}q^{2}}{%
(l^{2}+r_{+}^{2})(a^{2}+r_{+}^{2})}\right) },  \label{Mass}
\end{equation}%
where $\mathcal{W}(x)=LambertW(x)$. It is notable that $M$ has a minimum
which is corresponding to the extremal configuration for the black hole. So,
by taking the derivative of mass with respect to the horizon radius, one can
obtain the extremal charge which its maximum value is
\begin{equation}
q_{max}^{2}=\left( r_{+}^{2}-a^{2}+\dfrac{r_{+}^{2}(3r_{+}^{2}+a^{2})}{l^{2}}%
\right) \exp \left( \dfrac{3r_{+}^{4}+r_{+}^{2}(a^{2}+l^{2})-a^{2}l^{2}}{%
r_{+}^{4}+r_{+}^{2}(l^{2}+a^{2})+a^{2}l^{2}}\right) .  \label{qext}
\end{equation}

For the case of $M=l=1$, one can obtain the following relation between the
extremal value of quantities
\begin{equation}
q_{ext}=\sqrt{2r_{+}\left( 1+\dfrac{1-r_{+}^{2}}{1+r_{+}^{2}}\mathcal{W}%
\right) ,}
\end{equation}%
and
\begin{equation}
a_{ext}=\left( -1-\dfrac{2}{\mathcal{W}}\right) ^{1/2}r_{+},
\end{equation}%
where $\mathcal{W=W}\left( -r_{+}(r_{+}^{2}+1)exp\left( \dfrac{r_{+}^{2}-1}{%
r_{+}^{2}+1}\right) \right) .$

\begin{figure}[tbp]
\hspace{0.4cm} \centering
\subfigure[]{\includegraphics[width=0.4\columnwidth]{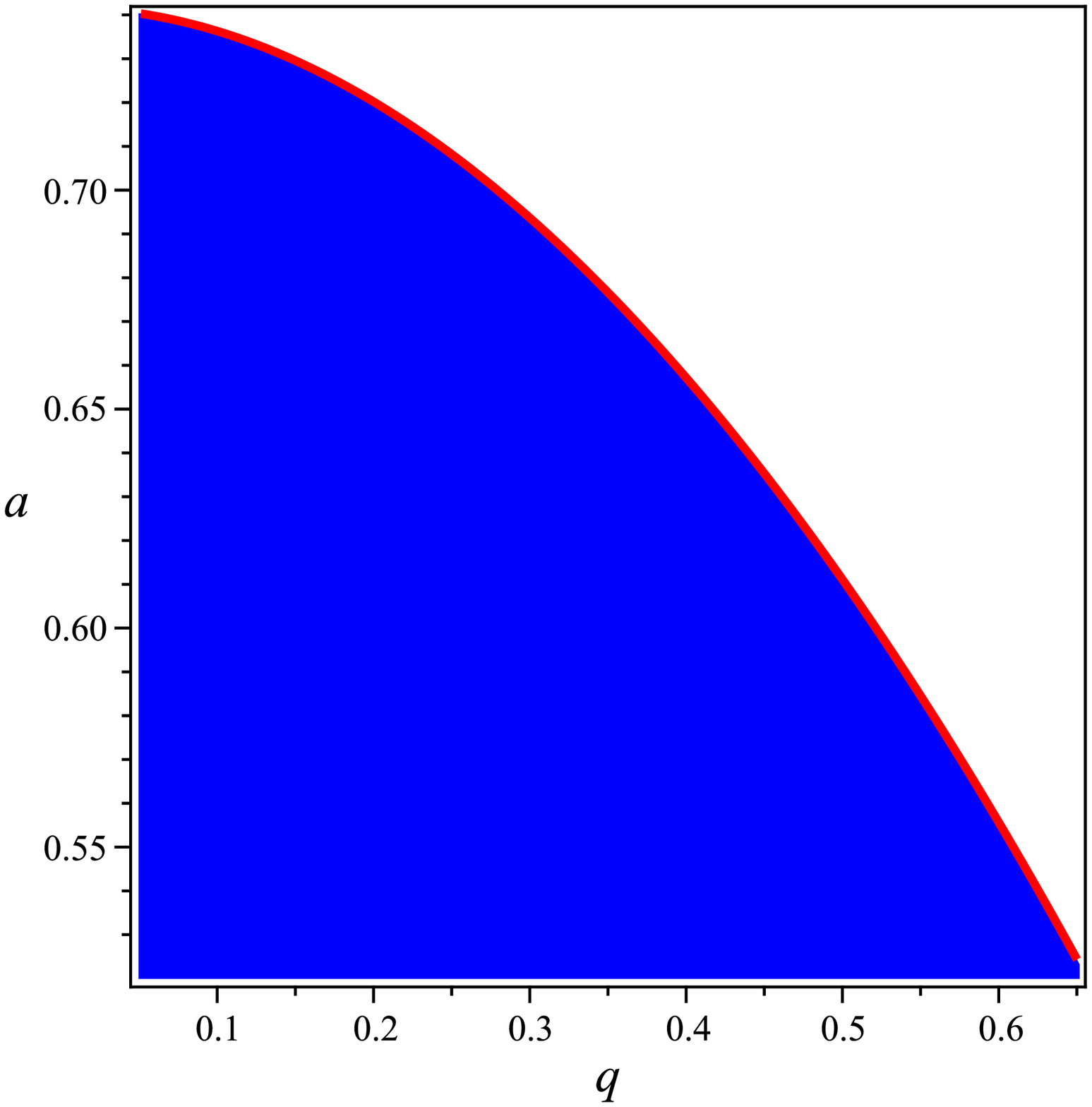}} %
\subfigure[]{\includegraphics[width=0.4\columnwidth]{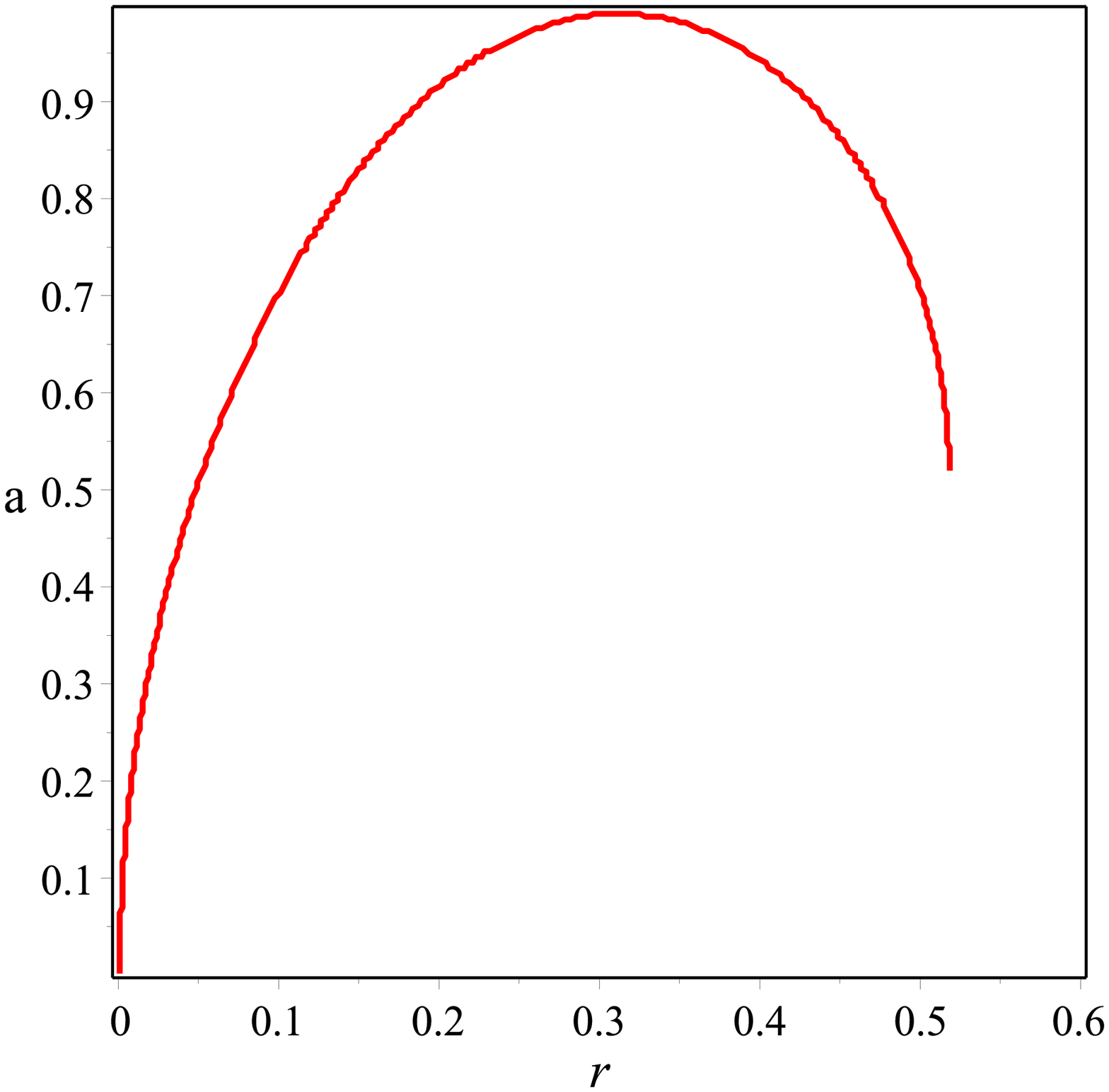}}
\caption{ The behavior of the critical value of the magnetic charge $q$ and
radius of the event horizon $r$ in terms of the rotation parameter $a$ for $%
M=1,l=1$. Notice that the $a-r$ plots are at the critical values of $q$. }
\label{figqaplot}
\end{figure}

The conditions of having the regular black hole or the no-horizon solution
in terms of the free parameters are observed in Fig. \ref{figqaplot}. In
order to have regular black holes, there are upper limits (critical values)
on the electric charge and rotation parameter of the metric. At the critical
values (the border of the shaded and white regions) there is a minimum
horizon which corresponds to the extremal black hole ($a-r$ plot). The
solution has no-horizon in the white region of the $a-q$ plot. In the case
of the rotating regular black hole (shaded region of $a-q$ plot), by
increasing the rotation parameter $a$ the critical value of the electric
charge decreases.
Static observers cannot exist everywhere in the spacetime, because the four
velocity of static observer finally becomes null. When this occurs the
observer cannot remain static and rotate with the black hole.Therefore, the
stationary limit surface is described by $g_{tt}=0$. Similar to the case of
event horizon, one can obtain the conditions for the critical parameter of
black hole so that the solutions of ($g_{tt}=0$) merge to one. The
conditions are%
\begin{equation}
g_{tt}=\partial _{r}g_{tt}=0.
\end{equation}%
\begin{figure}[tbp]
\hspace{0.4cm} \centering
\subfigure[]{\includegraphics[width=0.4\columnwidth]{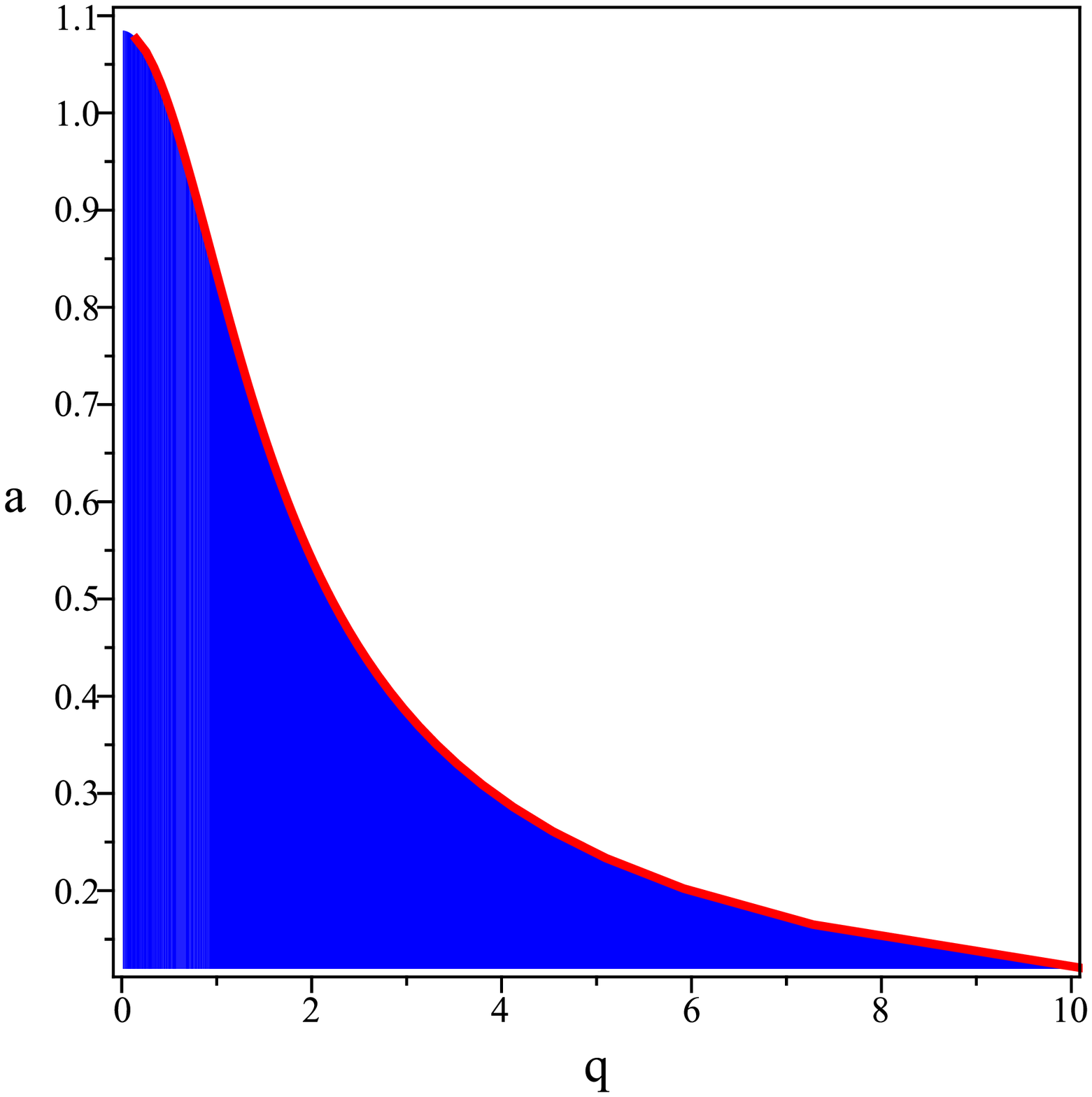}} %
\subfigure[]{\includegraphics[width=0.4\columnwidth]{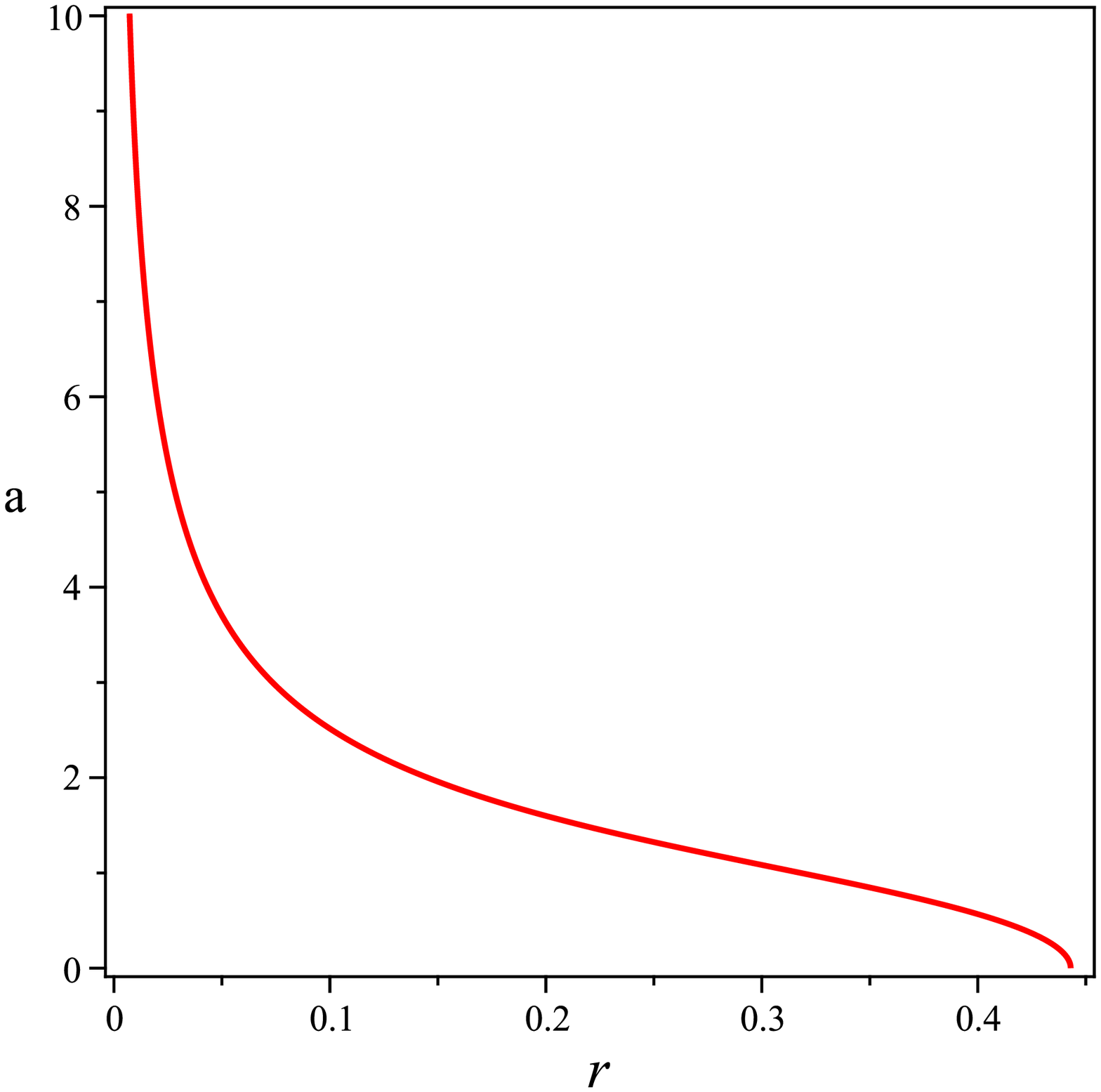}}
\caption{ The behavior of the critical value of the magnetic charge $q$ and
radius of the stationary limit surface $r$ in terms of the rotation
parameter $a$ for $M=1,l=1,\protect\theta =\protect\pi /2$. Notice that the $%
a-r$ plots are at the critical values of $q$. }
\label{figqaplotg}
\end{figure}
Solving the mentioned conditions, simultaneously, for obtaining $q$ and $r$
in the case of $M=l=1,\theta =\pi /2$ and plotting them, one can find Fig. %
\ref{figqaplotg} and obtain following equations
\begin{equation}
q=\sqrt{2r\left[ 1-\mathcal{W}(-er^{3})\right] },\hspace{0.5cm}a=\sqrt{%
-\left( 1+r^{2}+\dfrac{2r^{2}}{\mathcal{W}(-er^{3})}\right) }.
\end{equation}

In the $a-q$ plot, shaded and white plots correspond to the regular black
hole with stationary limit surface and without stationary limit surface,
respectively. The $a-r$ plot represents the dependence of the radius of the
stationary limit surface on $a$.
\begin{figure}[tbp]
\hspace{0.4cm} \centering
\subfigure[$a=0.3$]{\includegraphics[width=0.4\columnwidth]{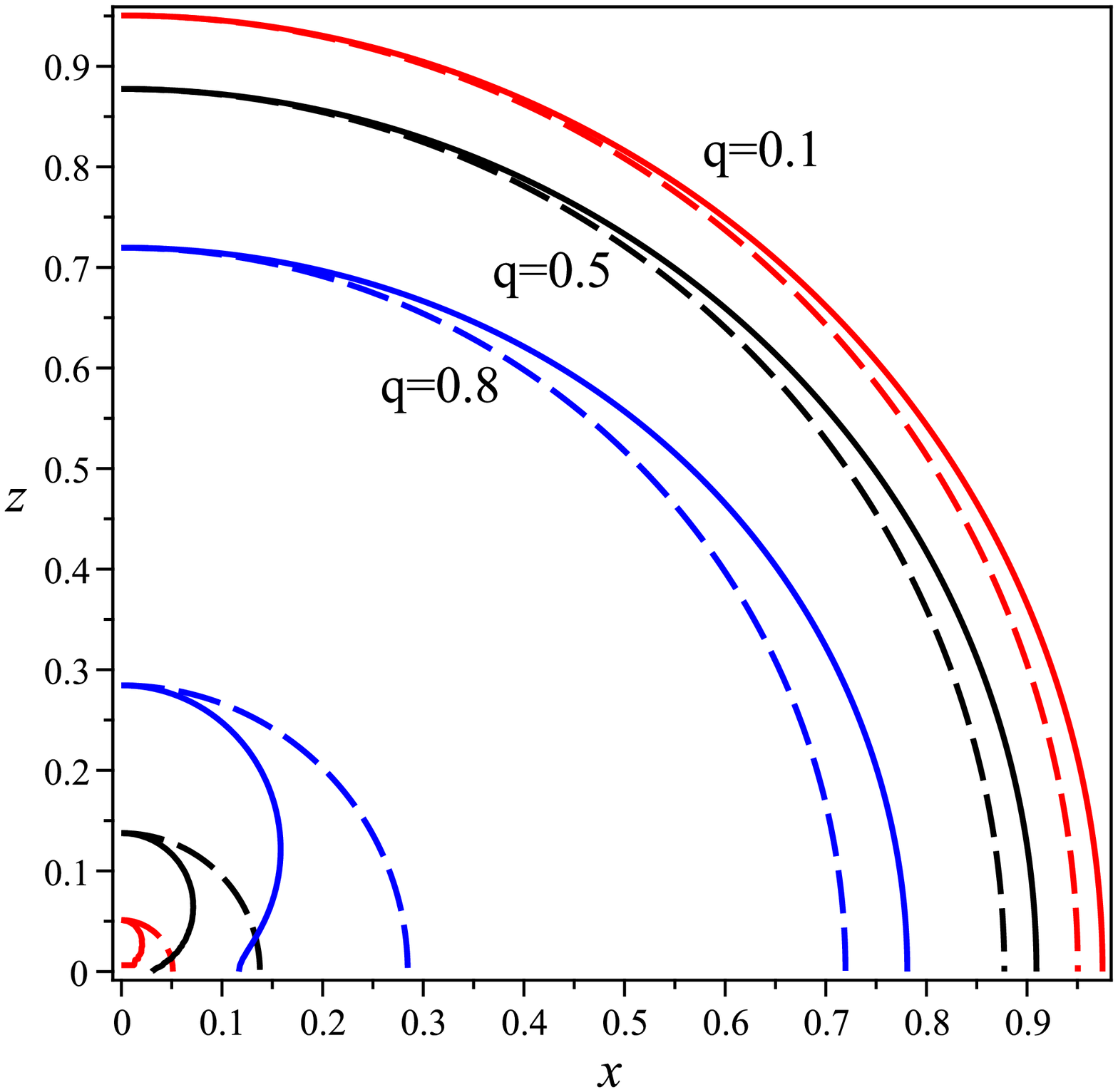}} %
\subfigure[$q=0.3$]{\includegraphics[width=0.4\columnwidth]{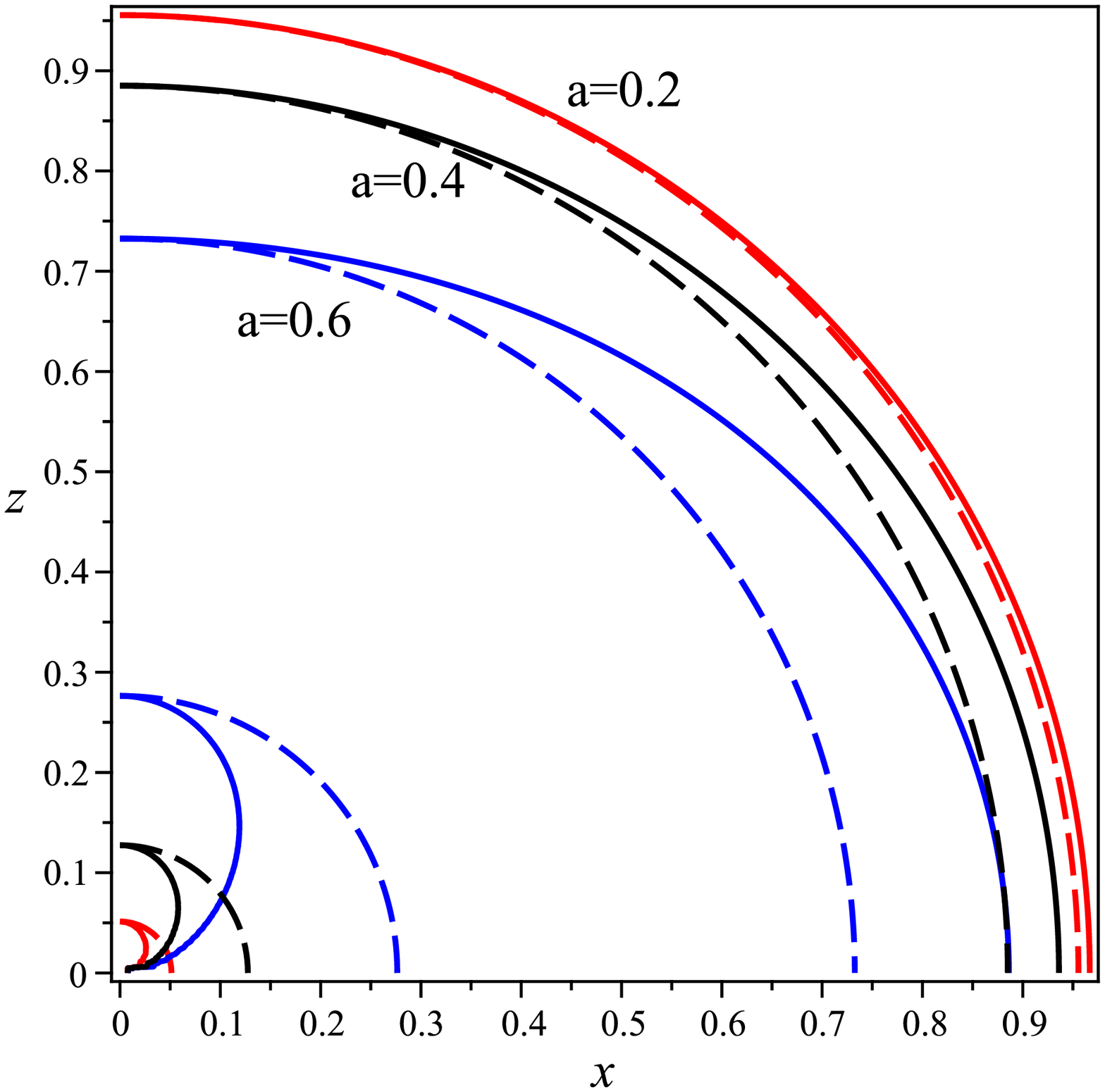}}
\caption{ Horizons (dashed line) and stationary limit surfaces (solid line)
for different values of $a$ and $q$ are depicted in the figures. }
\label{figzxplot}
\end{figure}
The stationary limit surface does not coincide with the event horizon and is
located outside the horizon. The region between the horizon and the
stationary limit surface is called the ergoregion which is shown in the Fig.
(\ref{figzxplot}). In Fig. (\ref{figzxplot}), size and shape of ergoregion
in the $z-x$ plane, where $z=r\cos (\theta )$ and $x=r\sin (\theta )$, have
been depicted. By increasing $q$ and $a$, one can observe the change in the
shape and size of the ergoregion.\newline
We now consider a possible nonlinear source for the metric (\ref{metric}).
The magnetic part of the gauge field ($A^{m}_{\mu}$) of charged rotating
regular black hole is given by%
\begin{eqnarray}
A_{\mu }^{m} &=&-\dfrac{qa\cos (\theta )}{\Sigma }\delta _{\mu }^{t}+\dfrac{%
q(r^{2}+a^{2})\cos (\theta )}{\Xi \Sigma }\delta _{\mu }^{\phi },
\end{eqnarray}%
in which by calculating the electromagnetic field tensor, one obtains%
\begin{equation}
\mathcal{F}=\mathcal{F_{\mu \nu }}\mathcal{F^{\mu \nu }}=\dfrac{%
2q^{2}(r^{4}-6r^{2}a^{2}\cos ^{2}(\theta )+a^{4}\cos ^{4}(\theta ))}{\Sigma
^{4}}.
\end{equation}%
By solving the Einstein tensor for $\mathcal{L}$ and $\mathcal{L_{F}}$ as
independent parameters, one can find%
\begin{equation}
\mathcal{L}=\dfrac{8r^{2}a^{2}\cos ^{2}(\theta )\Sigma f^{^{\prime \prime
}}+4\left( rf^{^{\prime }}-f\right) \left( r^{4}-6r^{2}a^{2}\cos ^{2}(\theta
)+a^{4}\cos ^{4}(\theta )\right) }{\Sigma ^{4}}-\dfrac{6}{l^{2}},
\label{eqla}
\end{equation}%
\begin{equation}
\mathcal{L_{F}}=\dfrac{-\Sigma f^{^{\prime \prime }}+4(rf^{^{\prime }}-f)}{%
2q^{2}}.  \label{eqlaf}
\end{equation}%
The above definitions for $\mathcal{L}$ and $\mathcal{L_{F}}$
satisfy all five different Einstein field equations. In the case
of $a=0$, one can recover the expressions presented in
\cite{Fan:2016hvf}. We should note that the total derivative of
$\mathcal{L}$ with respect to $\mathcal{F}$ is not equal to
$\mathcal{L_{F}}$ and their difference in the case of
$\theta=\theta_{0}=constant$ is given as
\begin{eqnarray}
\Delta\mathcal{L_{F}} &=&\mathcal{L_{F}}-\dfrac{\partial
\mathcal{L}}{\partial \mathcal{F}}=\mathcal{L_{F}}-\dfrac{\partial
\mathcal{L}}{\partial r}\dfrac{\partial r}{\partial
\mathcal{F}}\neq 0.
\end{eqnarray}
where its asymptotic limit ($r\gg 1$)is
\begin{eqnarray}
\Delta\mathcal{L_{F}} &\approx & -\dfrac{5 q^{2} a^{2}
\cos^{2}(\theta_{0})}{4 M r^3}+ \dfrac{3q^4 a^2
\cos^{2}(\theta_{0})}{4 M^2
r^4}+\mathcal{O}\left(\dfrac{1}{r^{5}}\right).
\end{eqnarray}

\begin{center}
\begin{figure}[]
\hspace{4cm}\includegraphics[width=8.cm]{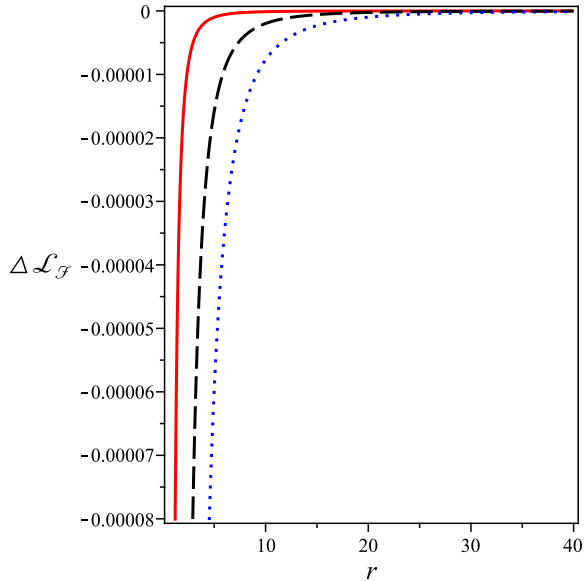}\vspace{0.1cm}
\caption{{\protect\small The behavior of $\Delta \mathcal{L_{F}}$ in terms
of $r$ at $\protect\theta=0$ for $M=1, a=0.1, q=0.1, 0.4, 0.8 $ (red to
blue).}}
\label{delt}
\end{figure}
\end{center}

In Fig. \ref{delt}, we have shown $\Delta \mathcal{L_{F}}$ in
terms of $r$ at the equatorial plane of the regular rotating black
holes for different values of parameters. According to the Fig.
(\ref{delt}), one finds that the inconsistency between
$\mathcal{L_{F}}$ and $\dfrac{\partial \mathcal{L}}{\partial
\mathcal{F}}$ is smaller than $10^{-4}$. As a result, we find that
the metric (\ref{metric}) is a charged rotating solution for the
Einstein equation.

\section{Thermodynamics \label{Therm}}

In this section, we explore the thermodynamics of the regular rotating-AdS
black hole solution (\ref{metric}). In order to investigate the
thermodynamic properties of the black hole in extended phase space, we need
to obtain some relevant thermodynamic quantities. In the extended phase
space, we treat the cosmological constant as a thermodynamic pressure and
its conjugate quantity as a thermodynamic volume via \cite{dolan10, dolan11,
altamirano14:1}%
\begin{equation}
P=\dfrac{3}{8\pi l^{2}},\hspace{1cm}V=\dfrac{r_{+}A}{3}+\dfrac{4\pi J^{2}}{3%
\mathcal{M}},  \label{eqvolum}
\end{equation}%
where $A$ is the horizon area of black hole which is calculated as%
\begin{equation}
A=\left. \int_{0}^{2\pi }\int_{0}^{\pi }\sqrt{g_{\theta \theta }g_{\phi \phi
}}d\theta d\phi \right\vert _{r_{+}}=\dfrac{4\pi (r_{+}^{2}+a^{2})}{\Xi }.
\label{eqentropy}
\end{equation}%
and $\mathcal{M}$ and $J$ are, respectively, the mass and the angular
momentum of the black hole which can be obtain by using of Altas-Tekin
method \cite{Altas:2018zjr}. Using the Killing vectors $k^{\mu }=\delta
_{t}^{\mu }/\Xi $ and $k^{\mu }=\delta _{\phi }^{\mu }$ associated with the
time translation and rotational invariance, one gets \cite{Caldarelli:1999xj}%
\begin{equation}
J=\dfrac{Ma}{\Xi ^{2}},\hspace{0.5cm}\mathcal{M}=\dfrac{M}{\Xi ^{2}}.
\label{eqmasang}
\end{equation}%
The Hawking temperature for non-extremal case can be obtained by using the
surface gravity interpretation%
\begin{eqnarray}
T &=&\dfrac{\kappa }{2\pi }=\left. \dfrac{1}{4\pi (a^{2}+r_{+}^{2})}\dfrac{%
d\Delta _{r}}{dr}\right\vert _{r_{+}}=-\dfrac{(2Mr_{+}+q^{2})e^{-\left(
\dfrac{q^{2}}{2Mr_{+}}\right) }}{4\pi (r_{+}^{2}+a^{2})r_{+}}+\dfrac{%
r_{+}(r_{+}^{2}+2r_{+}^{2}+a^{2})}{2\pi l^{2}(r_{+}^{2}+a^{2})},
\label{eqtem}
\end{eqnarray}%
where $\kappa $ is the surface gravity. In the case of small $q$ and $a$ we
have%
\begin{equation}
T =\dfrac{1}{4\pi r_{+}}\left( 1+\dfrac{3r_{+}^{2}}{l^{2}}-\dfrac{q^{2}}{%
r_{+}^{2}}\right) -\dfrac{1}{4\pi r_{+}^{3}}\left[ 2+\dfrac{2r_{+}^{2}}{l^{2}%
}-\dfrac{q^{2}}{r_{+}^{2}}\right] a^{2}+\mathcal{O}(q^{3},a^{4}).
\end{equation}%
%
%
%\begin{equation}
%T =\dfrac{1}{4\pi \text{r}_{+}}\left[ \dfrac{%
%3r_{+}^{4}+r_{+}^{2}l^{2}+a^{2}(r_{+}^{2}-l^{2})}{l^{2}(r_{+}^{2}+a^{2})}-%
%\dfrac{q^{2}}{r_{+}^{2}+a^{2}}\right] +\mathcal{O}(q^{3}).
%\end{equation}
As we know, the Killing vector $k^{\mu }=\delta _{t}^{\mu }+\Omega \delta
_{\phi }^{\mu }$ at the event horizon of the rotating black hole is a null
vector, and therefore, we can use%
\begin{equation}
k_{\mu }k^{\mu }=g_{tt}+2\Omega g_{t\phi }+\Omega ^{2}g_{\phi \phi }=0,
\end{equation}%
to obtain the angular velocity, $\Omega $, by inserting $g_{tt},g_{t\phi }$
and $g_{\phi \phi }$ from metric (\ref{metric})%
\begin{eqnarray}
\Omega &=&\dfrac{a\Xi ((r_{+}^{2}+a^{2})\Delta _{\theta }-\Delta _{r})}{%
\Delta _{\theta }(r_{+}^{2}+a^{2})^{2}-\Delta _{r}a^{2}\sin ^{2}(\theta )}%
\pm \dfrac{\Xi \Sigma \sqrt{\Delta _{r}\Delta _{\theta }}}{(\Delta
_{r}a^{2}\sin ^{2}(\theta )-\Delta _{\theta }(r_{+}^{2}+a^{2})^{2})\sin
(\theta )},
\end{eqnarray}%
on the horizon $\Delta _{r}(r_{+})=0$, so we obtain following expression for
angular velocity
\begin{equation}
\Omega _{+}=\dfrac{a\Xi }{r_{+}^{2}+a^{2}}.  \label{eqomega}
\end{equation}%
However, we should note that the thermodynamical angular velocity is the
differences between the angular velocity measured by the observer at the
infinity and the angular velocity at the horizon, yielding%
\begin{equation}
\Omega =\Omega _{+}-\Omega _{\infty }=\Omega _{+}+\dfrac{a^{2}}{l^{2}}=%
\dfrac{a}{r_{+}^{2}+a^{2}}\left( 1+\dfrac{r_{+}^{2}}{l^{2}}\right) .
\end{equation}%
The electric part of vector potential is given as%
\begin{equation}
A_{\mu }^{e}=-\dfrac{qr}{\Sigma }\delta _{\mu }^{t}+\dfrac{qar\sin
^{2}(\theta )}{\Sigma \Xi }\delta _{\mu }^{\phi },
\end{equation}%
so, the electrostatic potential of the event horizon with respect to spatial
infinity as an electrostatic potential reference is obtained as%
\begin{equation}
\Phi =\left. A_{\mu }^{e}k^{\mu }\right\vert _{\infty }-\left. A_{\mu
}^{e}k^{\mu }\right\vert _{r_{+}}=\dfrac{qr_{+}}{r_{+}^{2}+a^{2}},
\label{eqphi}
\end{equation}%
where $k^{\mu }=\delta _{t}^{\mu }+\Omega _{+}\delta _{\phi }^{\mu }$ is the
null generator of the horizon. By computing the flux of electromagnetic
field tensor by using of the standard Gauss' law, one can obtain the
electric charge as follow%
\begin{equation}
Q=\dfrac{1}{4\pi }\int \star F=\dfrac{q}{\Xi },  \label{eqcharge}
\end{equation}%
where $\star F$ is the dual of the Faraday 2-form. For a black hole embedded
in AdS spacetime, employing the relation between the cosmological constant
and thermodynamic pressure, would result in the interpretation of the black
hole mass as the enthalpy. Using the expressions (\ref{eqmasang}), (\ref%
{eqcharge}) and (\ref{eqentropy}) for mass, angular momentum, electric
charge and entropy, and the fact that $\Delta _{r}(r_{+})=0$, one obtains
the enthalpy in terms of thermodynamic quantities as%
\begin{equation}
H=\mathcal{M}=\left( \dfrac{8\pi PJ^{2}}{3}+\dfrac{\pi J^{2}}{S}+\dfrac{\pi
Q^{4}}{S\Upsilon ^{2}}\right) ^{\frac{1}{2}},\hspace{0.5cm}\Upsilon =%
\mathcal{W}\left( \dfrac{3\pi Q^{2}}{8PS(S+\dfrac{3}{8P})}\right)
\label{eqentalpy}
\end{equation}%
by using (\ref{eqentalpy}), one can determine the temperature, electrostatic
potential, volume and angular momentum respectively, as%
\begin{equation}
T=\left( \dfrac{\partial H}{\partial S}\right) _{Q,P,J}=-\dfrac{\pi J^{2}}{2%
\mathcal{M}S^{2}}-\dfrac{\pi Q^{4}\left( \Upsilon (PS+3/8)-(3SP+3/8)\right)
}{\mathcal{M}S^{2}\Upsilon ^{2}(1+\Upsilon )(3+8PS)},  \label{temp2}
\end{equation}%
\begin{equation}
\Phi =\left( \dfrac{\partial H}{\partial Q}\right) _{S,P,J}=\dfrac{\pi Q^{3}%
}{2\mathcal{M}S\Upsilon (1+\Upsilon )},  \label{eqphi11}
\end{equation}%
\begin{equation}
V=\left( \dfrac{\partial H}{\partial P}\right) _{Q,S,J}=\dfrac{4\pi J^{2}}{3%
\mathcal{M}}+\dfrac{2\pi Q^{4}}{\mathcal{M}(3+8PS)\Upsilon ^{2}(1+\Upsilon )}%
,
\end{equation}%
\begin{equation}
\Omega =\left( \dfrac{\partial H}{\partial J}\right) _{Q,P,S}=\dfrac{\pi
J(8PS+3)}{3\mathcal{M}S}.  \label{omega2}
\end{equation}%
Calculations show that the intensive quantities calculated by Eqs. (\ref%
{temp2})-(\ref{omega2}) coincide with Eqs. (\ref{eqvolum}), (\ref{eqtem}), (%
\ref{eqomega}) and (\ref{eqphi}) respectively. For instance in Fig. (\ref%
{phii}), we have done a comparison between relation (\ref{eqphi11}) (%
\textcolor{red}{red solid line}) and (\ref{eqphi}) (%
\textcolor{blue}{blue
solid line}).

\begin{center}
\begin{figure}[]
\hspace{4cm}\includegraphics[width=8.cm]{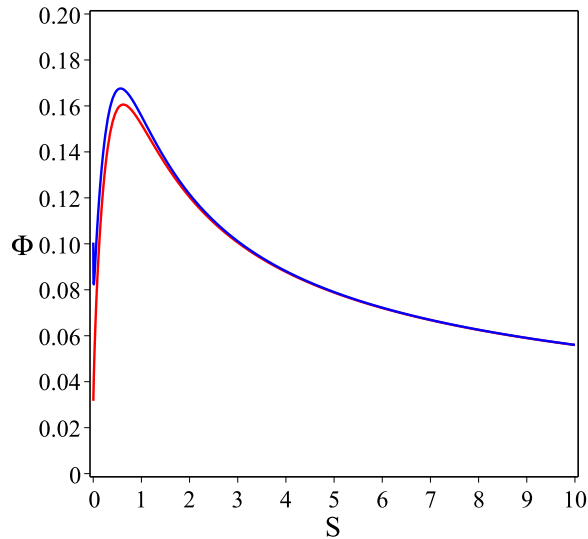}\vspace{0.1cm}
\caption{{\protect\small The behavior of $\Phi$ in terms of $S$ for typical
values of $Q=J=P=0.1$.}}
\label{phii}
\end{figure}
\end{center}

\begin{center}
\begin{figure}[tbp]
\hspace{4cm}\includegraphics[width=8.cm]{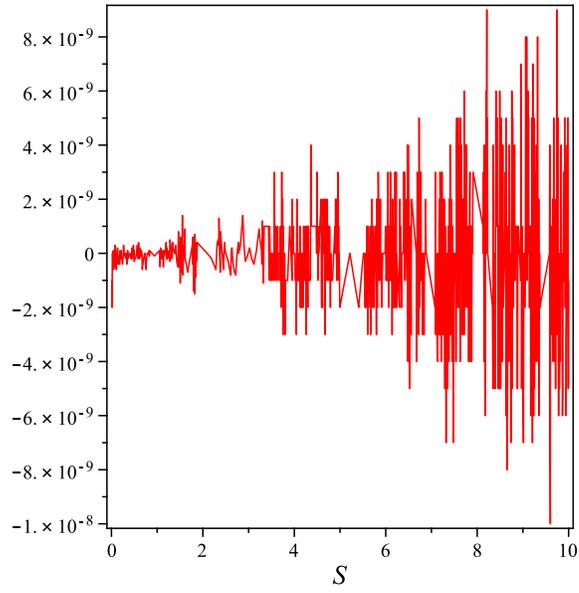}\vspace{0.1cm}
\caption{{\protect\small The differences between left and right of the Smarr
relation in terms of $S$ for typical values of $Q=J=P=0.1$.}}
\label{dif}
\end{figure}
\end{center}

Thus, these thermodynamic quantities satisfy the first law of black hole
thermodynamics in the enthalpy representation
\begin{equation}
dH=TdS+\Phi dQ+VdP+\Omega dJ.
\end{equation}%
In addition, for the sake of completeness, we calculate the Smarr relation.
Using the scaling argument, it should be given as%
\begin{equation}
H=2TS+Q\Phi +2\Omega J-2PV.  \label{eqsmarr}
\end{equation}%
In figure \ref{dif}, we have shown the differences between left and right of
equation (\ref{eqsmarr}). As can be seen, the figure shows the familiar
Smarr relation is satisfied.

\subsection{Phase transition and stability}

\begin{figure}[]
\centering
\subfigure[]{
 \includegraphics[width=0.3\columnwidth]{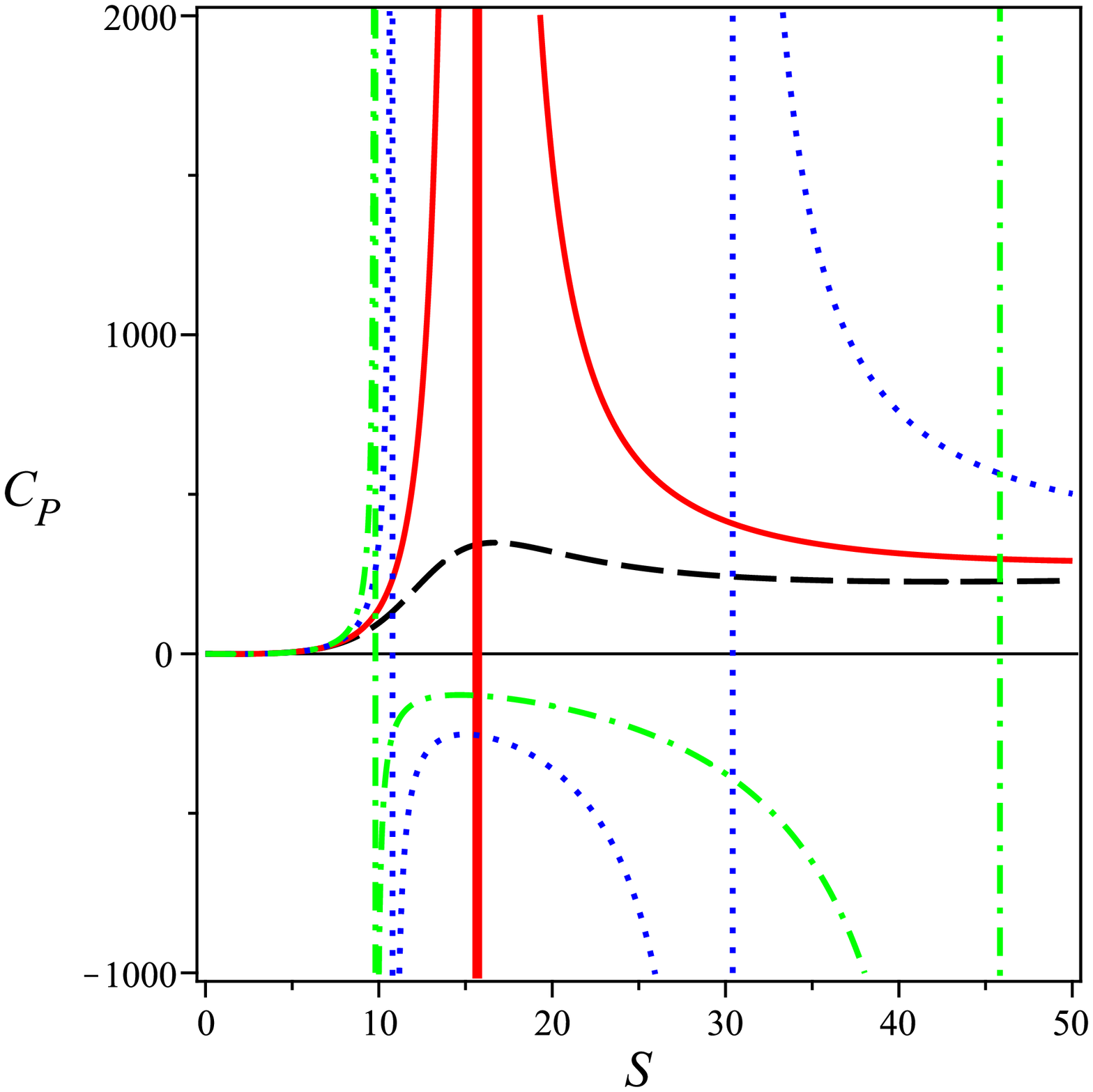}
 \label{figCP1}
 }
\subfigure[]  {\  \includegraphics[width=0.3\columnwidth]{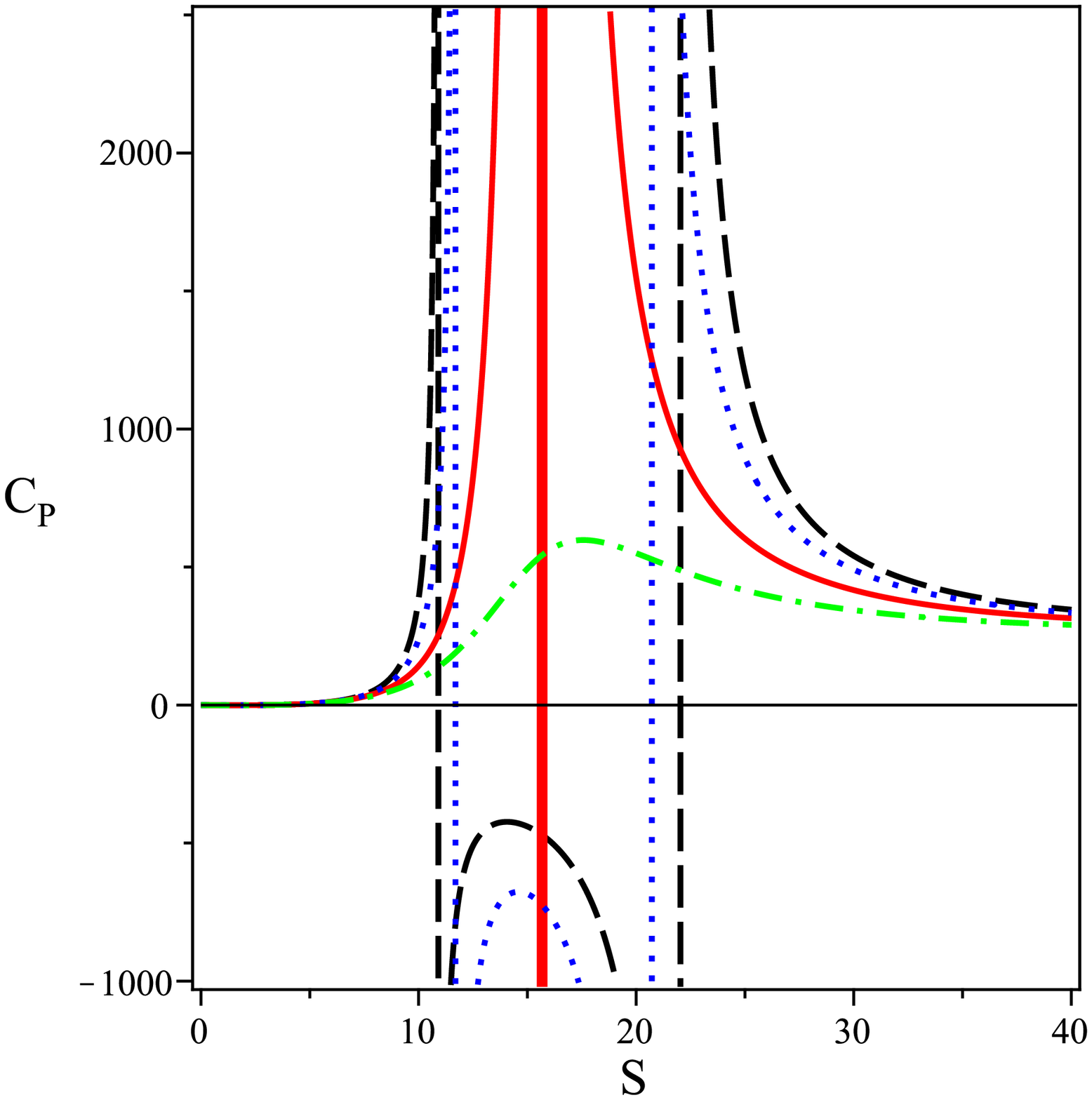}
\label{figCP2}  }
\subfigure[]  {\
\includegraphics[width=0.3\columnwidth]{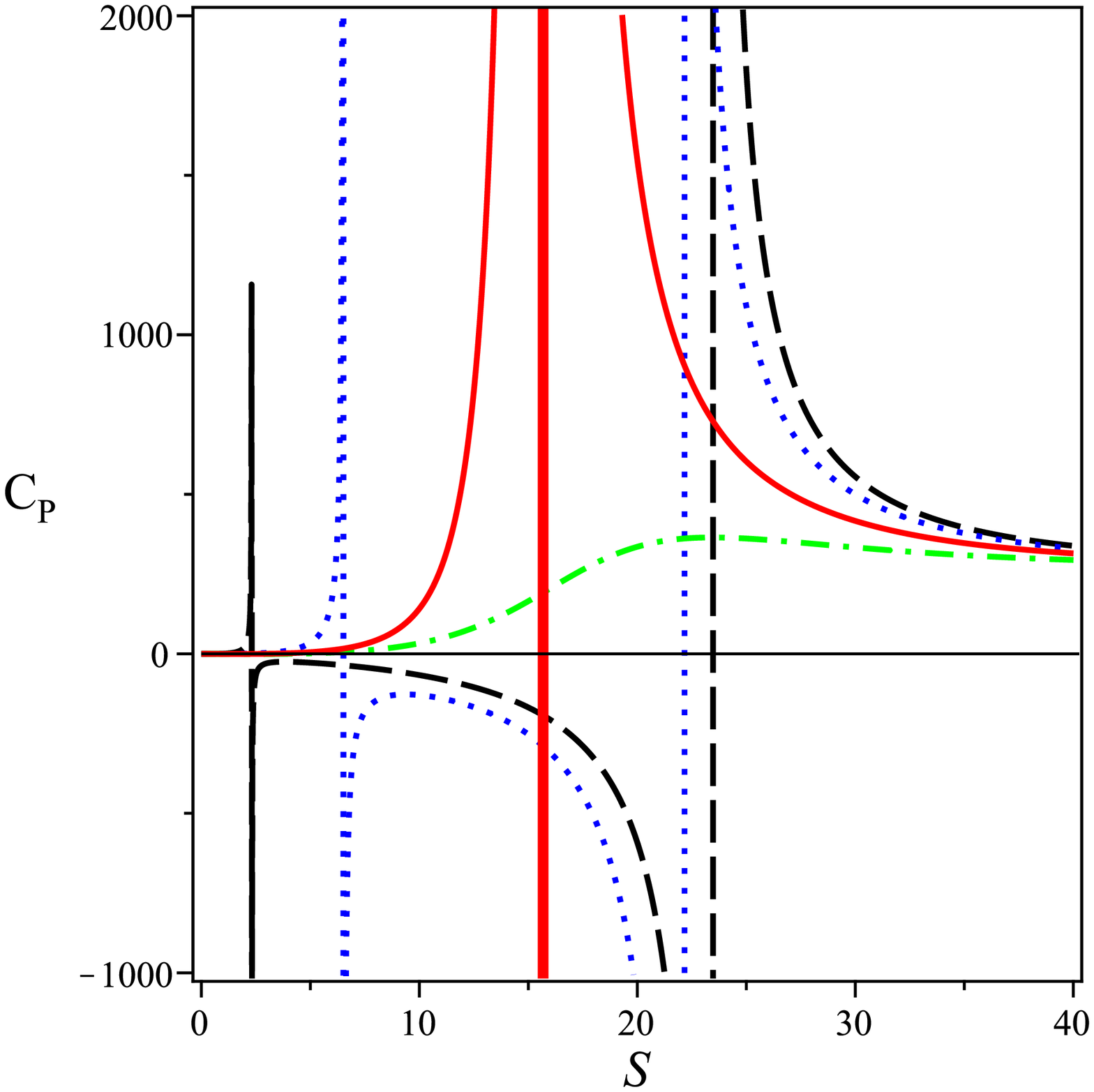}  \label{figCP3}  }
\caption{The behavior of $C_{P}$ in terms of $S $ for $J=0.5, Q=0.5$, $P=
\textcolor{green}{0.0025}<P_{crit} $, $P=\textcolor{blue}{0.0035}<P_{crit} $%
, $P=P_{crit}=\textcolor{red}{0.004802}$ and $P=\textcolor{black}{0.006}%
>P_{crit}$ (left), for $P=0.004802$, $J=0.5$ and $Q=\textcolor{black}{0.1},
\textcolor{blue}{0.3},\textcolor{red}{0.5},\textcolor{green}{0.7}$ (middle),
for $P=0.004802$, $Q=0.5$ and $J=\textcolor{black}{0.1},\textcolor{blue}{0.3}%
,\textcolor{red}{0.5},\textcolor{green}{0.7}$ (right).}
\label{figCPplot}
\end{figure}

Thermodynamic stability tells us how a system in thermodynamic equilibrium
responds to fluctuations of thermodynamic parameters. We should distinguish
between global and local stability. In global stability, we allow a system
in equilibrium with a thermodynamic reservoir to exchange energy with the
reservoir. The preferred phase of the system is the one that minimizes the
Gibbs free energy. In order to investigate the global stability, we use the
following expression for the Gibbs free energy in terms of $S$, $Q$ and $J$
\begin{equation}
G=H-TS=\dfrac{4J^{2}(128S^{2}P^{2}+120SP+27)+\frac{3Q^{4}[3\left( 3\Upsilon
+1\right) +8SP\left( 3\Upsilon -1\right) ]}{(1+\Upsilon )\Upsilon ^{2}}}{%
4S(8SP+3)\left[ \frac{3}{\pi S}\left( 32SJ^{2}P+12J^{2}+\dfrac{3Q^{4}}{%
\Upsilon ^{2}}\right) \right] ^{1/2}},
\end{equation}%
in which at the large $S$, we have%
\begin{equation}
G=\dfrac{1}{4}\sqrt{\dfrac{S}{\pi }}\left( 1-\dfrac{8SP}{3}+\dfrac{3\pi Q^{2}%
}{S}+\dfrac{10\pi ^{2}J^{2}}{S^{2}}\right) +\mathcal{O}\left( \dfrac{1}{S^{4}%
}\right) .
\end{equation}%
\begin{figure}[]
\centering
\subfigure[]{
 \includegraphics[width=0.3\columnwidth]{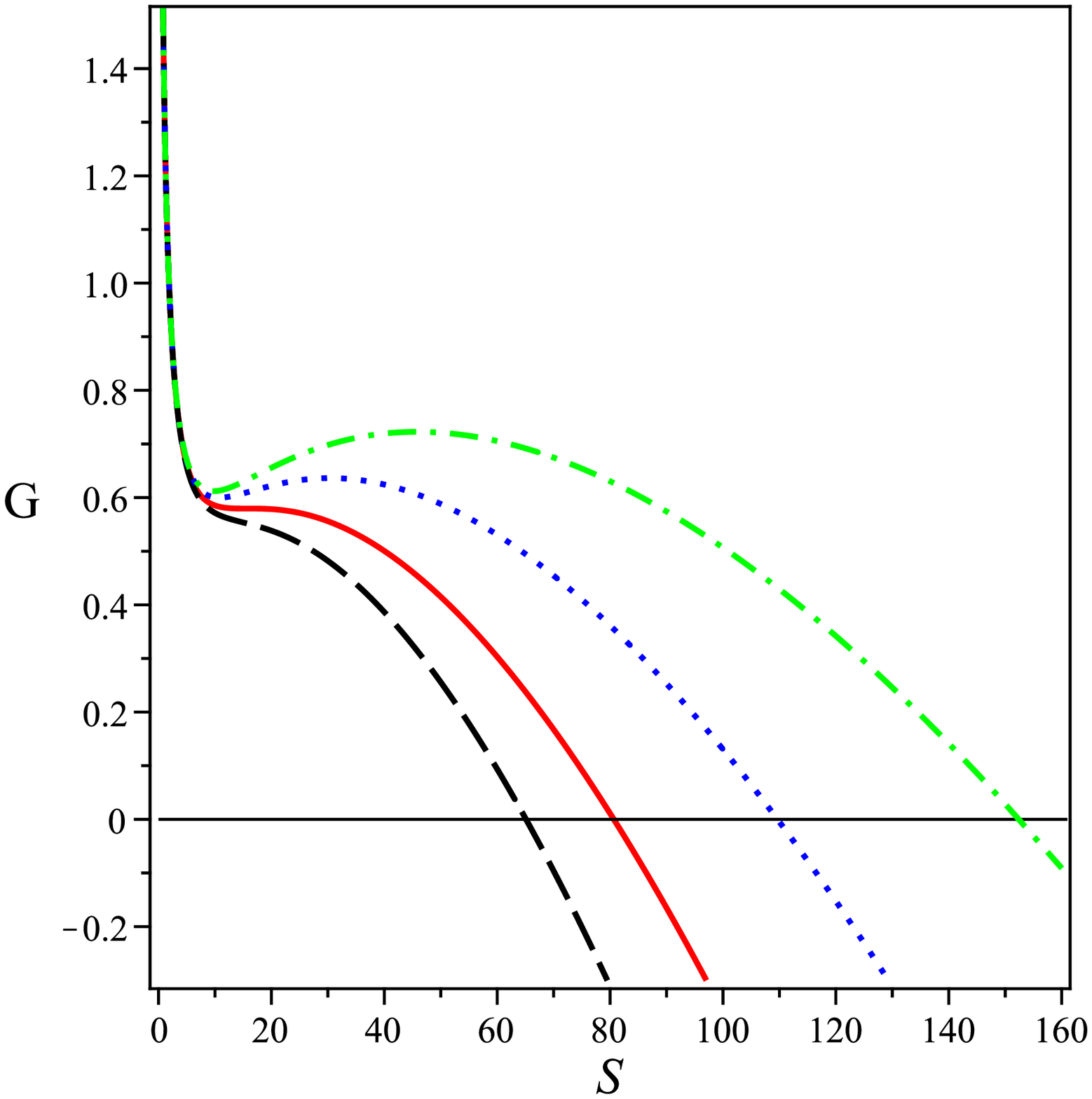}
 \label{fig:FDMsavingCW1}
 }
\subfigure[]  {\  \includegraphics[width=0.3\columnwidth]{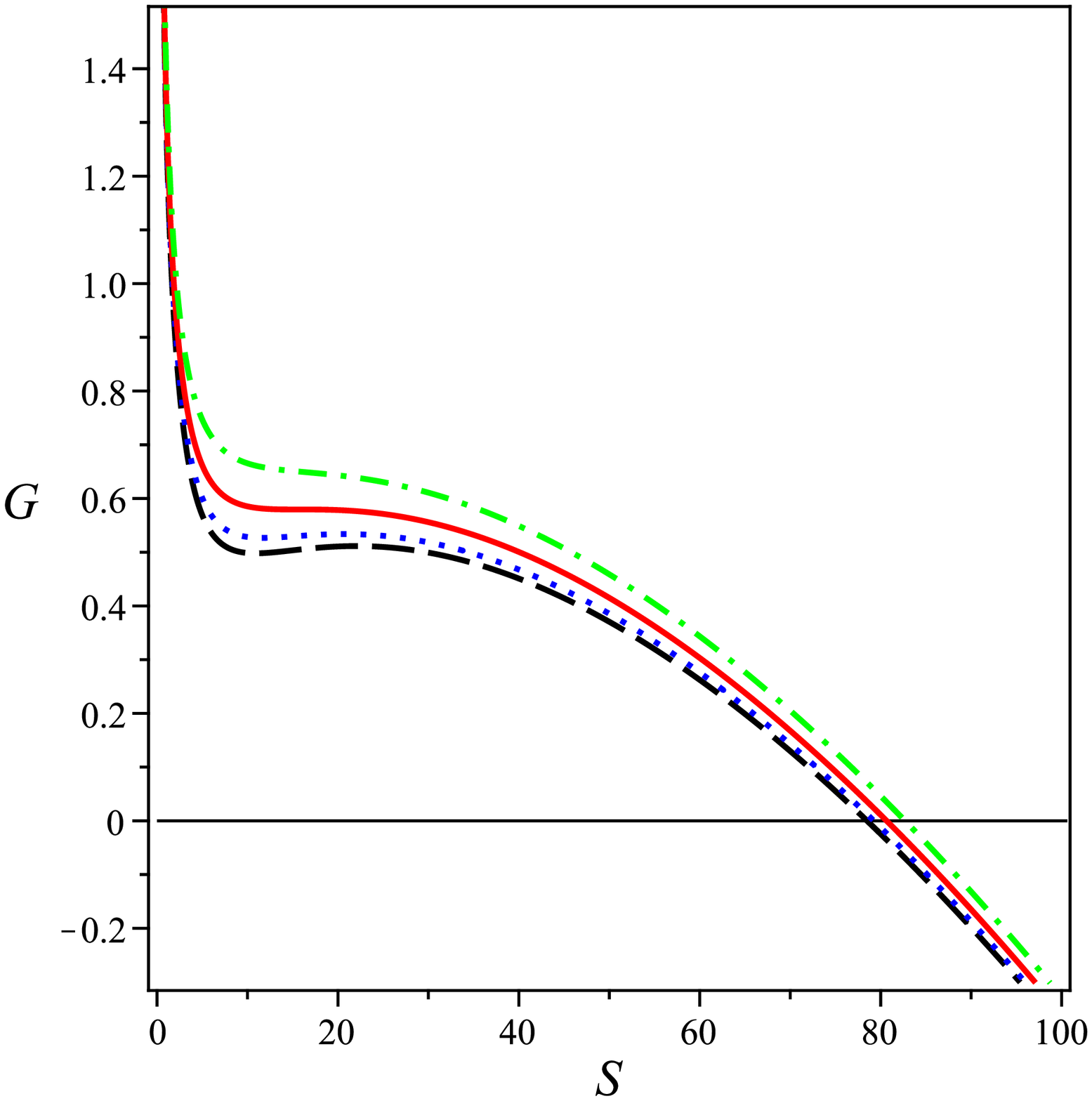}
\label{fig:FDMsavingCW2}  }
\subfigure[]  {\
\includegraphics[width=0.3\columnwidth]{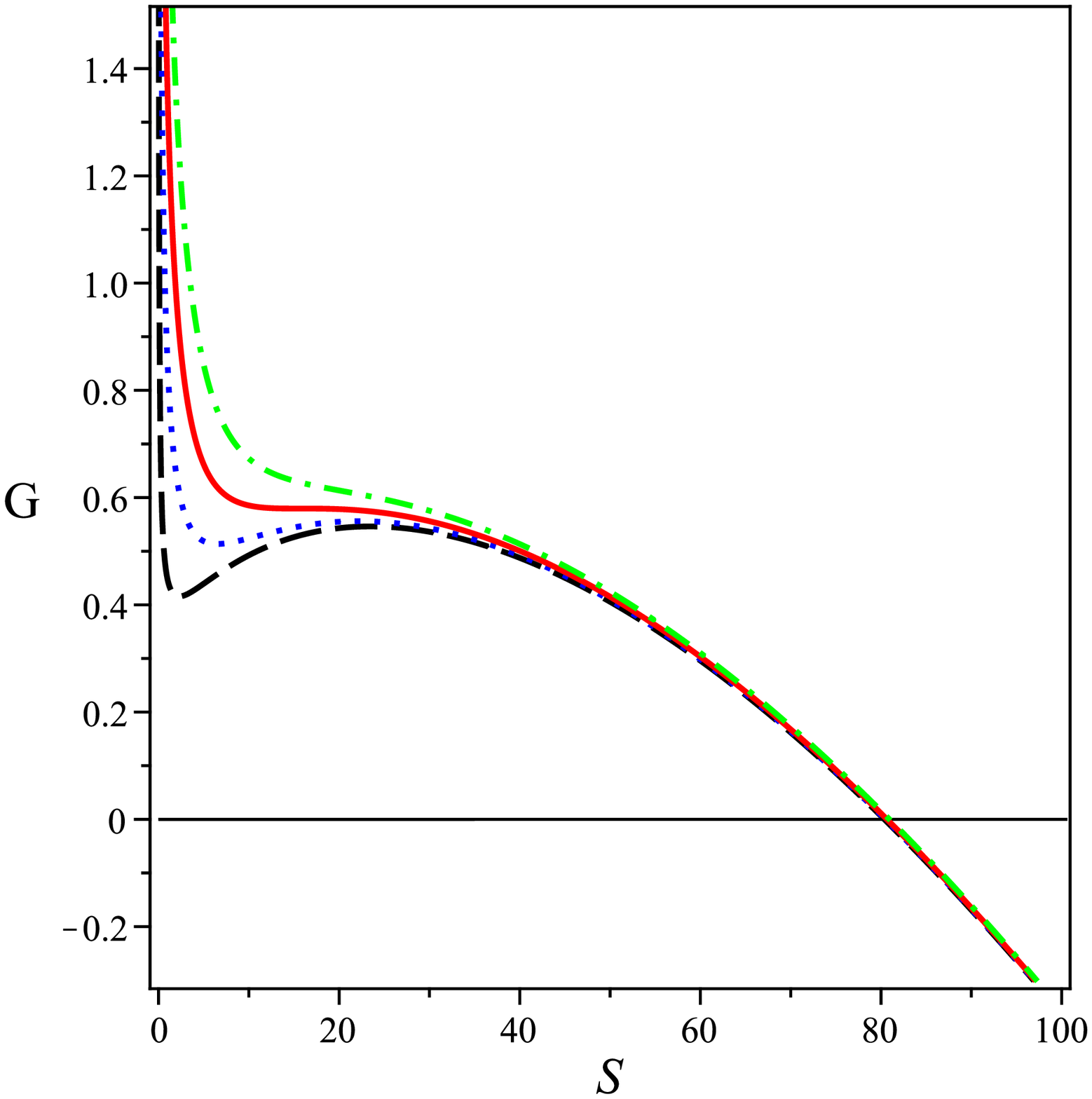}  \label{fig:FDMsavingCW3}  }
\caption{The behavior of $G$ in terms of $S$ for $P=\textcolor{green}{0.0025}%
<P_{crit}$, $P=\textcolor{blue}{0.0035}<P_{crit}$, $P=P_{crit}=
\textcolor{red}{0.004802}$, $P=\textcolor{black}{0.006}>P_{crit}$ and $%
J=0.5,Q=0.5$ (left), for $P=0.004802$, $J=0.5$ and $Q=\textcolor{black}{0.1}%
, \textcolor{blue}{0.3},\textcolor{red}{0.5},\textcolor{green}{0.7}$
(middle), for $P=0.004802$, $Q=0.5$ and $J=\textcolor{black}{0.1},
\textcolor{blue}{0.3},\textcolor{red}{0.5},\textcolor{green}{0.7}$ (right).}
\label{figgrplot}
\end{figure}
In Fig. (\ref{figgrplot}), we have shown the Gibbs free energy in terms of
entropy. Considering Fig. (\ref{fig:FDMsavingCW1}), we find that for
constant $J$ and $Q$, the Gibbs free energy is a decreasing function of $S$
for both small and large event horizon entropy, while it is an increasing
function for intermediate $S$. This behavior confirms that intermediate
black holes are globally unstable. Since the large black holes have negative
Gibbs free energy, they are more stable than small black holes. Also, Fig. (%
\ref{figgrplot}) shows that by increasing the pressure the black hole is
more stable. Also, we have shown the behavior of the Gibbs free energy in
terms of $S$ for constant pressure and angular momentum in Fig. (\ref%
{fig:FDMsavingCW2}) and for constant pressure and electric charge in Fig. (%
\ref{fig:FDMsavingCW3}). It is obvious that by increasing the angular
momentum and electric charge the black hole is more stable.\newline
On the other hand, local stability is concerned with how the system responds
to small changes in its thermodynamic parameters. In order to study the
thermodynamic stability of the black holes with respect to small variations
of the thermodynamic coordinates, one can investigate the behavior of the
heat capacity. The positivity of the heat capacity ensures the local
stability. The form of $C_{P}$, plotted in Fig. (\ref{figCPplot}), is
explicitly as follow%
\begin{equation}
C_{P}=T\left( \dfrac{\partial S}{\partial T}\right) _{P}=\dfrac{%
-2S(1+\Upsilon )^{2}\Gamma _{3}(4J^{2}\Upsilon ^{2}\Gamma _{3}+3Q^{4})\Theta
_{1}}{\Theta _{2}}  \label{CP}
\end{equation}%
where
\begin{equation}
\Theta _{1}=\Upsilon \Gamma _{3}[Q^{4}+4J^{2}\Upsilon (1+\Upsilon
)]-3Q^{4}\Gamma _{1}
\end{equation}%
and
\begin{align}
\Theta _{2}& =64J^{4}\Upsilon ^{5}\Gamma _{3}^{2}\Gamma _{9/4}\left[
\Upsilon ^{2}+3(\Upsilon +1)\right] +4096J^{4}\Upsilon ^{4}\Gamma _{3}\left(
S^{2}P^{2}+\dfrac{21}{32}SP+\dfrac{27}{256}\right)  \notag \\
& +16Q^{4}J^{2}\Upsilon ^{5}\Gamma _{3}^{2}\Gamma
_{9/2}-3072Q^{4}J^{2}\Upsilon ^{4}\Gamma _{3}\left( S^{2}P^{2}-\dfrac{3}{16}%
SP-\dfrac{15}{128}\right) +18Q^{8}\Gamma _{-3/2}  \notag \\
& +3072J^{2}Q^{4}\Upsilon ^{3}\Gamma _{3}\left( S^{2}P^{2}+\dfrac{15}{16}SP+%
\dfrac{21}{128}\right) +9Q^{8}\Upsilon ^{3}\Gamma
_{3}^{2}+48J^{2}Q^{4}\Upsilon ^{2}\Gamma _{3}^{2}\Gamma _{3/2}  \notag \\
& -1344Q^{8}\Upsilon ^{2}\left( S^{2}P^{2}+\dfrac{3}{28}SP-\dfrac{9}{448}%
\right) +27Q^{8}\Upsilon (64S^{2}P^{2}+\dfrac{20}{3}SP+1).
\end{align}%
in which we used the abbreviation of $\Gamma _{i}=i+8SP$.

As can be seen from Fig. (\ref{figCP1}), there exist three different regions
for $P<P_{crit}$. The partially positive specific heat for small black hole
region and large black hole region means that those black holes are
thermodynamically locally stable. Having negative specific heat of the
intermediate black hole region represents a locally unstable system. The
unstable region disappears at pressure $P=P_{crit}$ resulting in a
divergence point. When $P>P_{crit}$, $C_{P}$ is always positive and no
divergent point exists. This means that in this case the black hole is local
stable for arbitrary values of $S$. In Fig. (\ref{figCP2}) and (\ref{figCP3}%
), the behavior of $C_{P}$ in terms of $S$ for different values of electric
charge $Q$ and angular momentum $J$ of black hole has been plotted.
Obviously, at the critical pressure by increasing $Q$ and $J$ the critical
point disappears and black hole is stable.

\begin{figure}[]
\centering
\subfigure[]{
 \includegraphics[width=0.3\columnwidth]{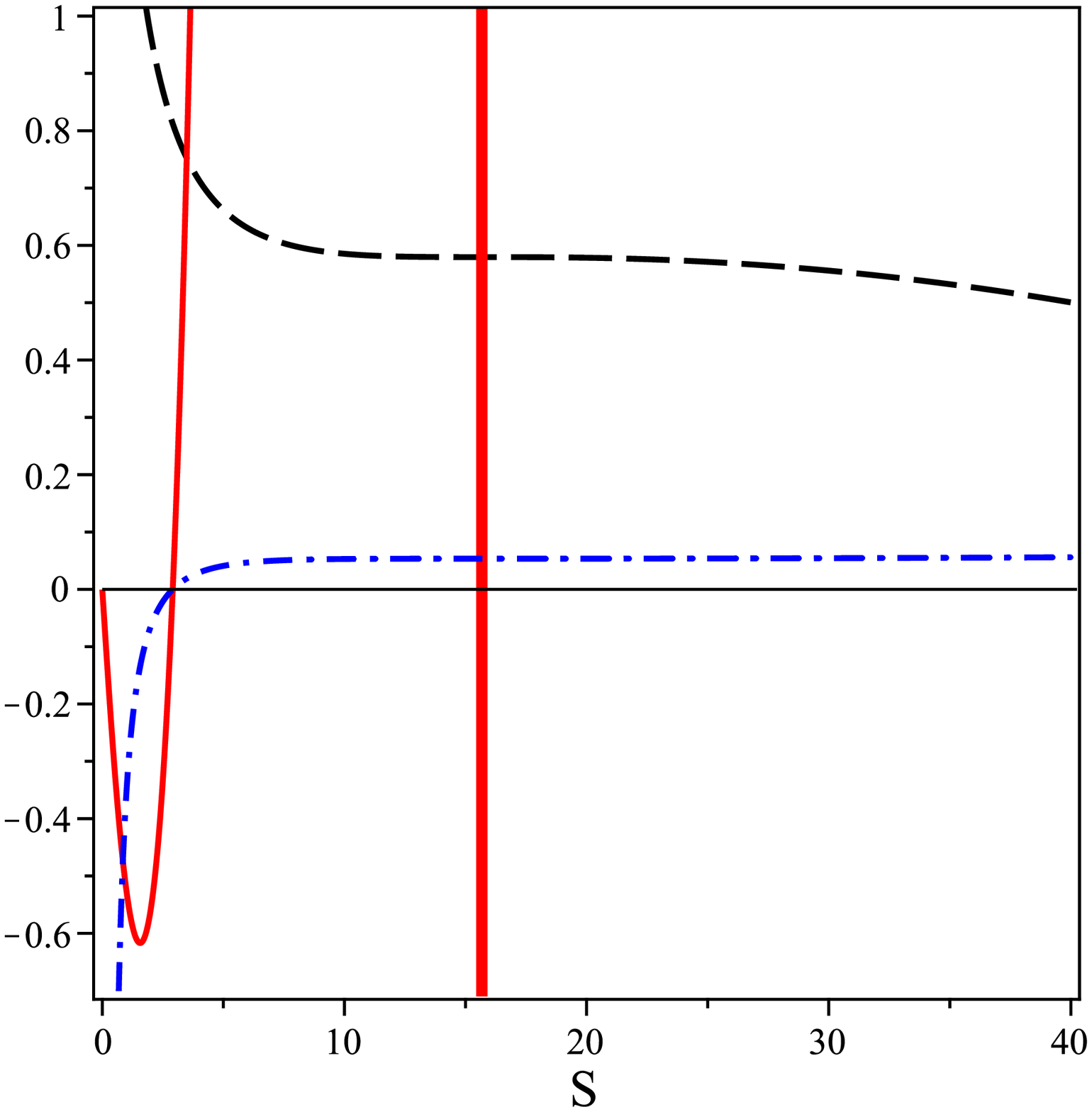}
 \label{figCG1}
 }
\subfigure[]  {\  \includegraphics[width=0.3\columnwidth]{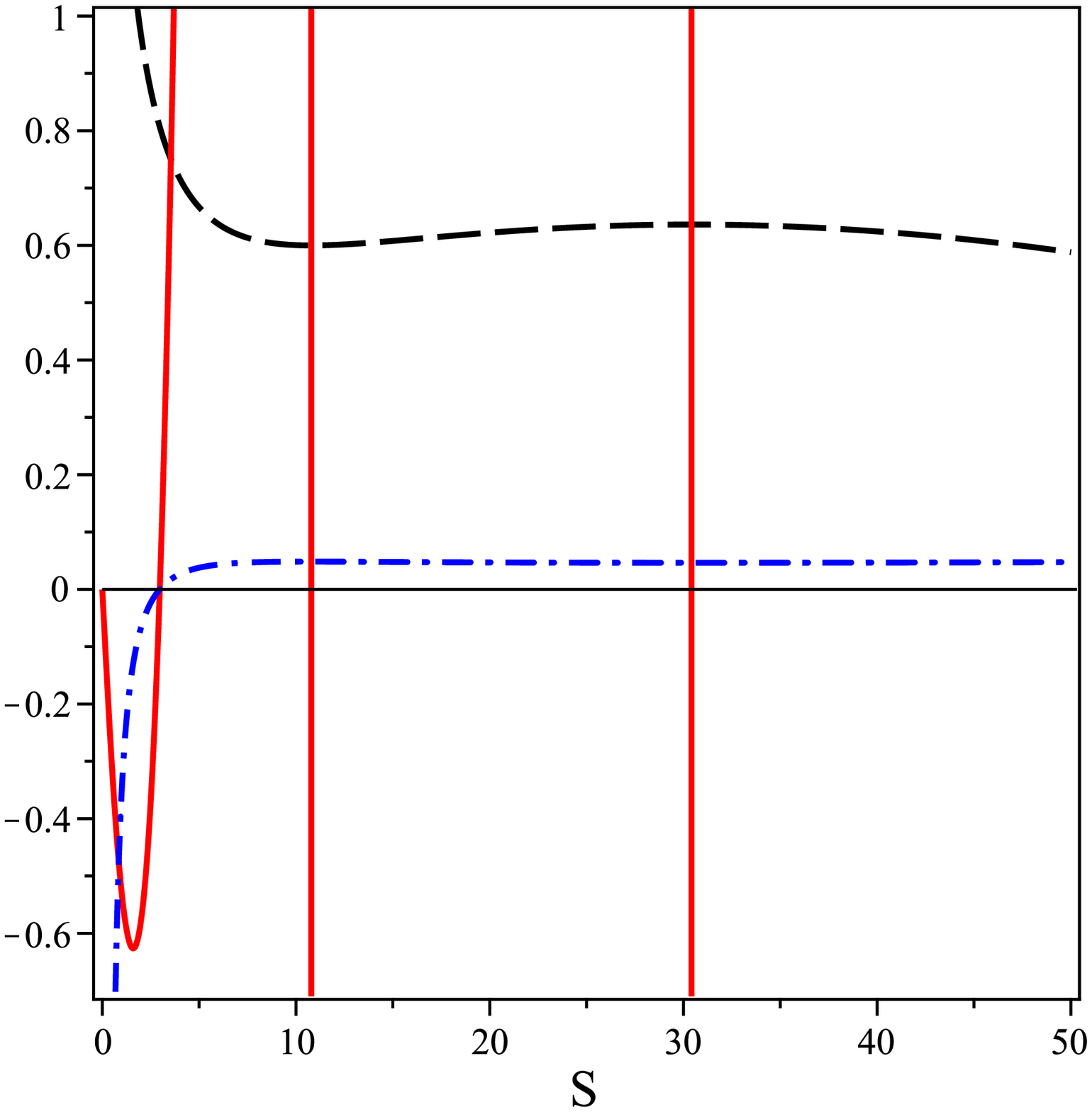}
\label{figCG2}  }
\subfigure[]  {\
\includegraphics[width=0.3\columnwidth]{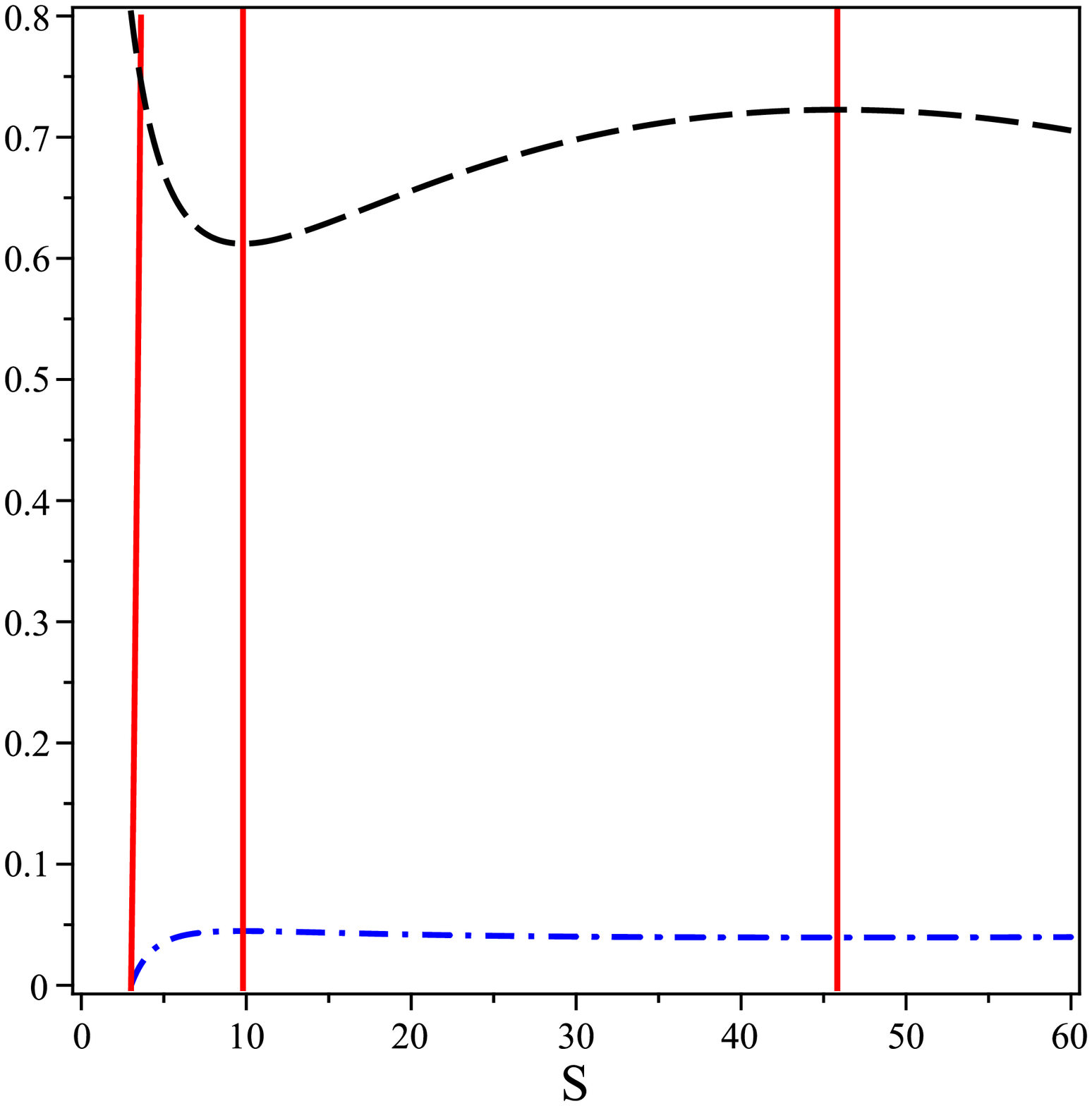}  \label{figCG3}  }
\caption{The behavior of $C_P$ (red solid line), $T $ ( blue dotted line)
and $G$ ( black dashed line) in terms of $S $ for $J=0.5, Q=0.5$, $%
P=0.004802 $ (left), $P=0.0035$ (middle) and $P=0.0025$ (right).}
\label{CGplot}
\end{figure}

\begin{figure}[]
\centering
\subfigure[]{
 \includegraphics[width=0.3\columnwidth]{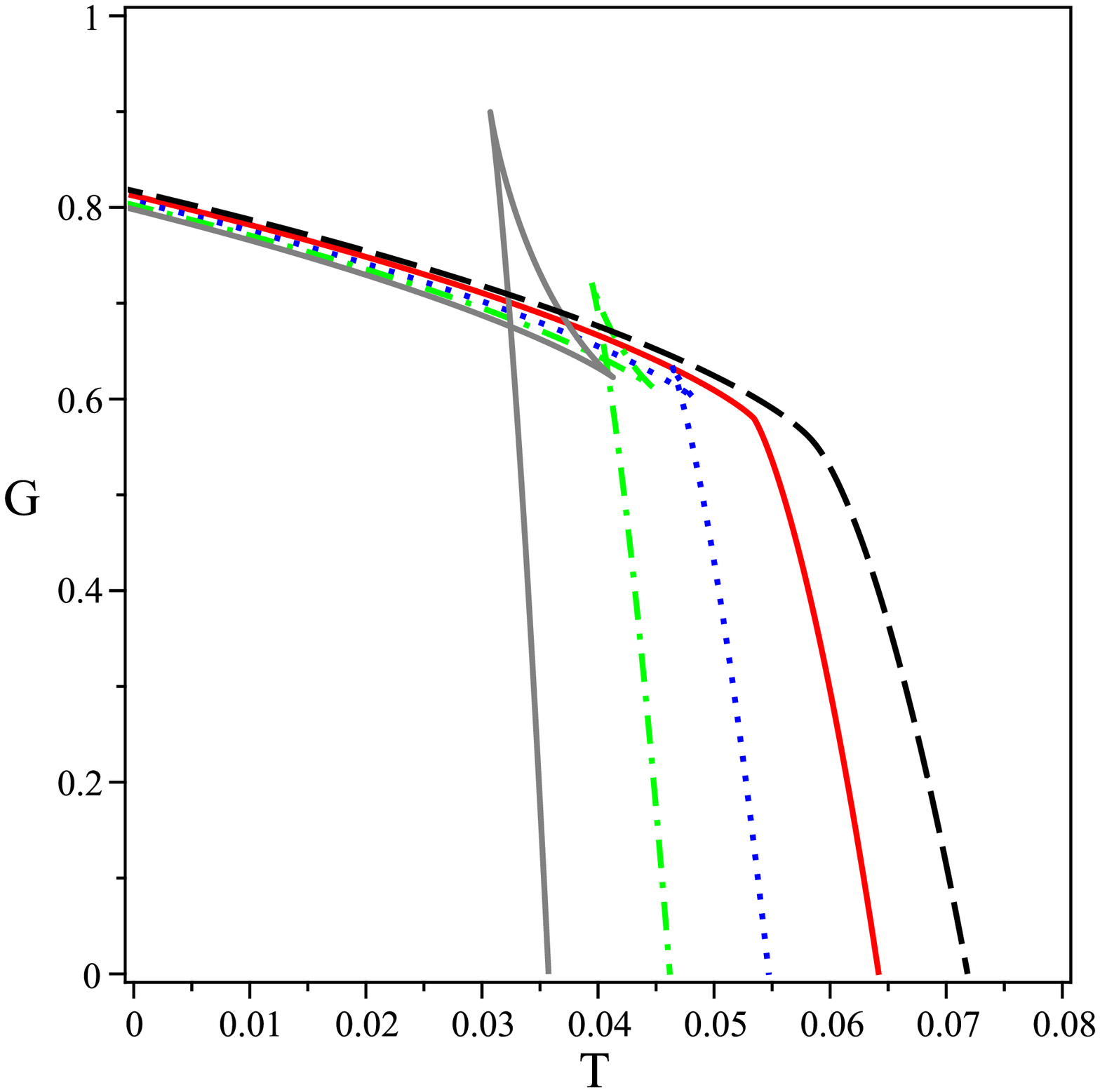}
 \label{figGT1}
 }
\subfigure[]  {\  \includegraphics[width=0.3\columnwidth]{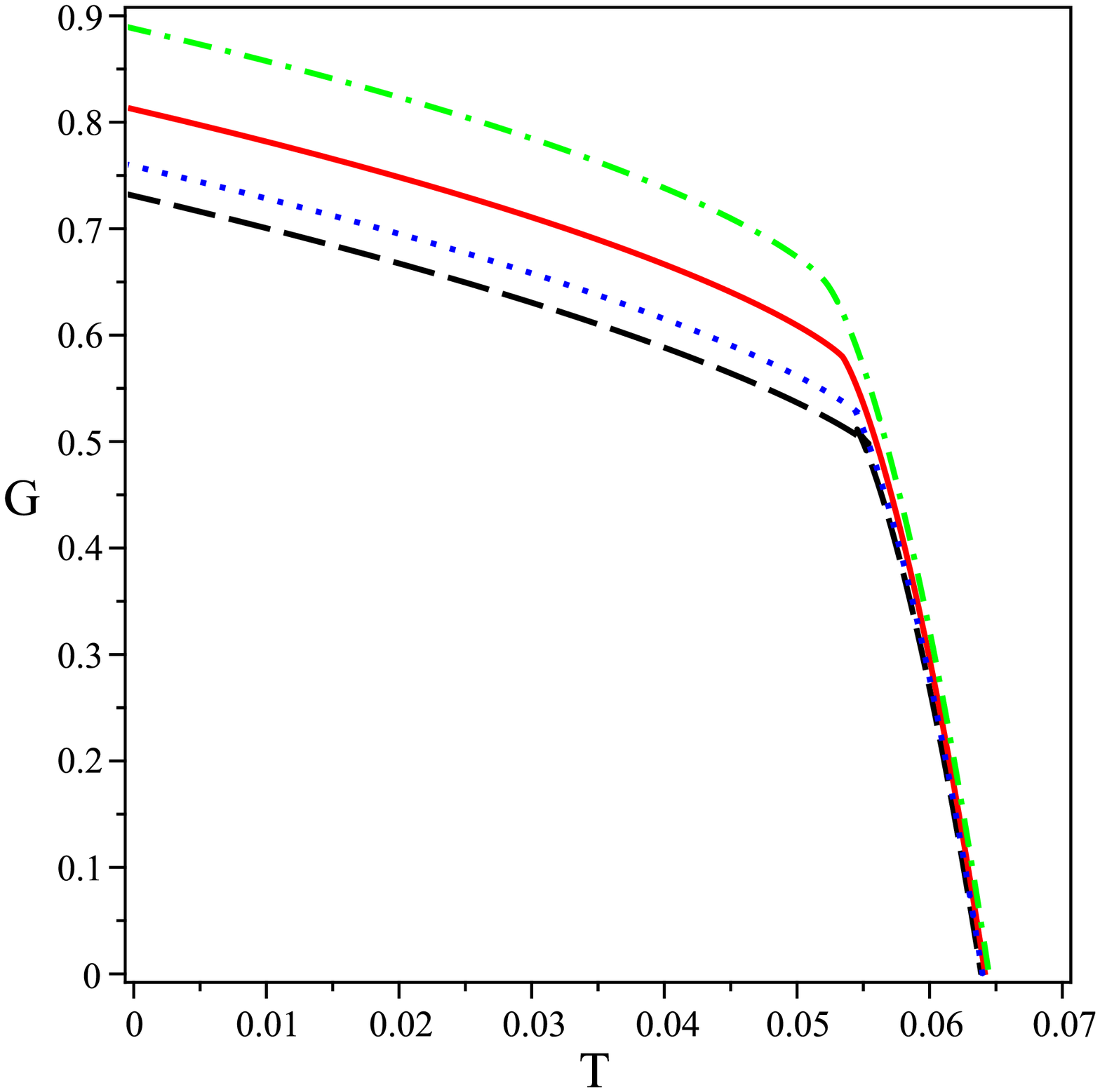}
\label{figGT2}  }
\subfigure[]  {\
\includegraphics[width=0.3\columnwidth]{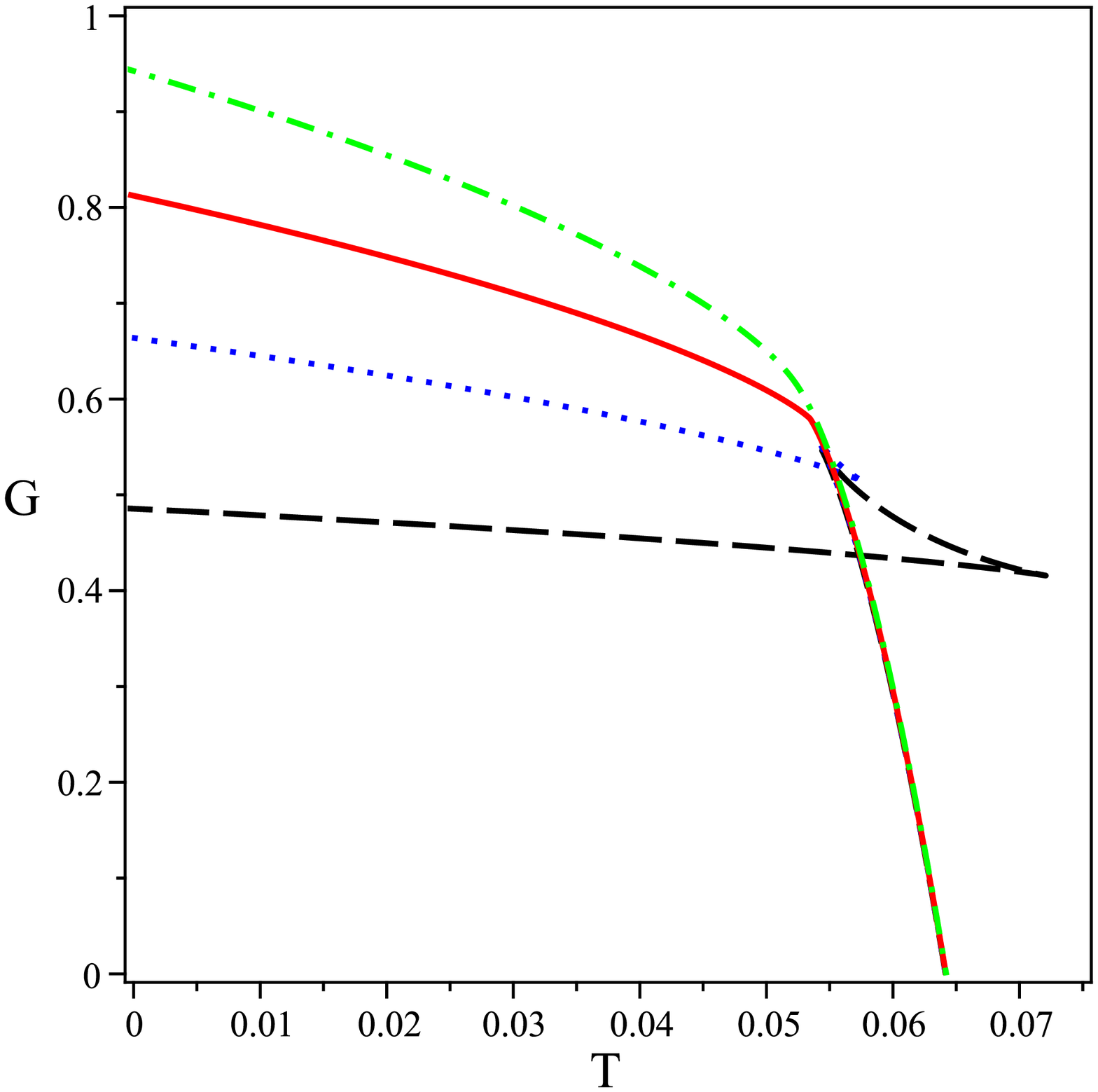}  \label{figGT3}  }
\caption{The behavior of $G$ in terms of $T $ for $P=0.0015<P_{crit} $, $%
P=0.0025<P_{crit} $, $P=\textcolor{green}{0.0035}<P_{crit} $, $P=P_{crit}=
\textcolor{red}{0.004802}$, $P=\textcolor{blue}{0.006}>P_{crit}$ and $J=0.5,
Q=0.5$ (left), for $P=0.004802$, $J=0.5$ and $Q=\textcolor{black}{0.1},
\textcolor{blue}{0.3},\textcolor{red}{0.5},\textcolor{green}{0.7}$ (middle),
for $P=0.004802$, $Q=0.5$ and $J=\textcolor{black}{0.1},\textcolor{blue}{0.3}%
,\textcolor{red}{0.5},\textcolor{green}{0.7}$ (right).}
\label{GTplot}
\end{figure}

The behavior of heat capacity could also represent the phase transitions. It
is clear that the heat capacity has at most two divergences which separate
small stable BHs, unstable region with medium horizon radius, and large
stable ones that are coincidence with extremum points of temperature and
Gibbs free energy. Also, the root of temperature coincides with the point
that the heat capacity changes its signature. The root separates the
non-physical solutions with negative temperature from physical black holes
with positive temperature (Fig. \ref{CGplot}).\newline
At the root the temperature is zero in which corresponds to the extremal
configurations of black hole and leads to the maximum of the angular
momentum such that
\begin{equation}
J_{max}^2=\dfrac{(3(1-\Upsilon)+8SP(3-\Upsilon))Q^4}{4\Upsilon^2(1+%
\Upsilon)(3+8SP)}.
\end{equation}
The plot of the Gibbs free energy with respect to the temperature shows a
swallowtail behavior as presented in Fig. (\ref{GTplot}). When $P < %
\textcolor{red}{0.004802}$, the Gibbs free energy with respect to
temperature develops a swallowtail like shape. There is a small/large first
order phase transition in the black hole, which resembles the liquid/gas
change of phase occurring in the van der Waals fluid. At the critical
pressure $P=\textcolor{red}{0.004802}$ , the swallowtail disappears which
corresponds to the critical point. In the Fig. (\ref{figGT2}) and (\ref%
{figGT3}), we have represented the behavior of the Gibbs free energy in
terms temperature for the different values of $Q$ and $J$. As can be seen,
by increasing $J$ and $Q$ the critical point disappears.

\begin{figure}[]
\centering
\subfigure[]{
 \includegraphics[width=0.3\columnwidth]{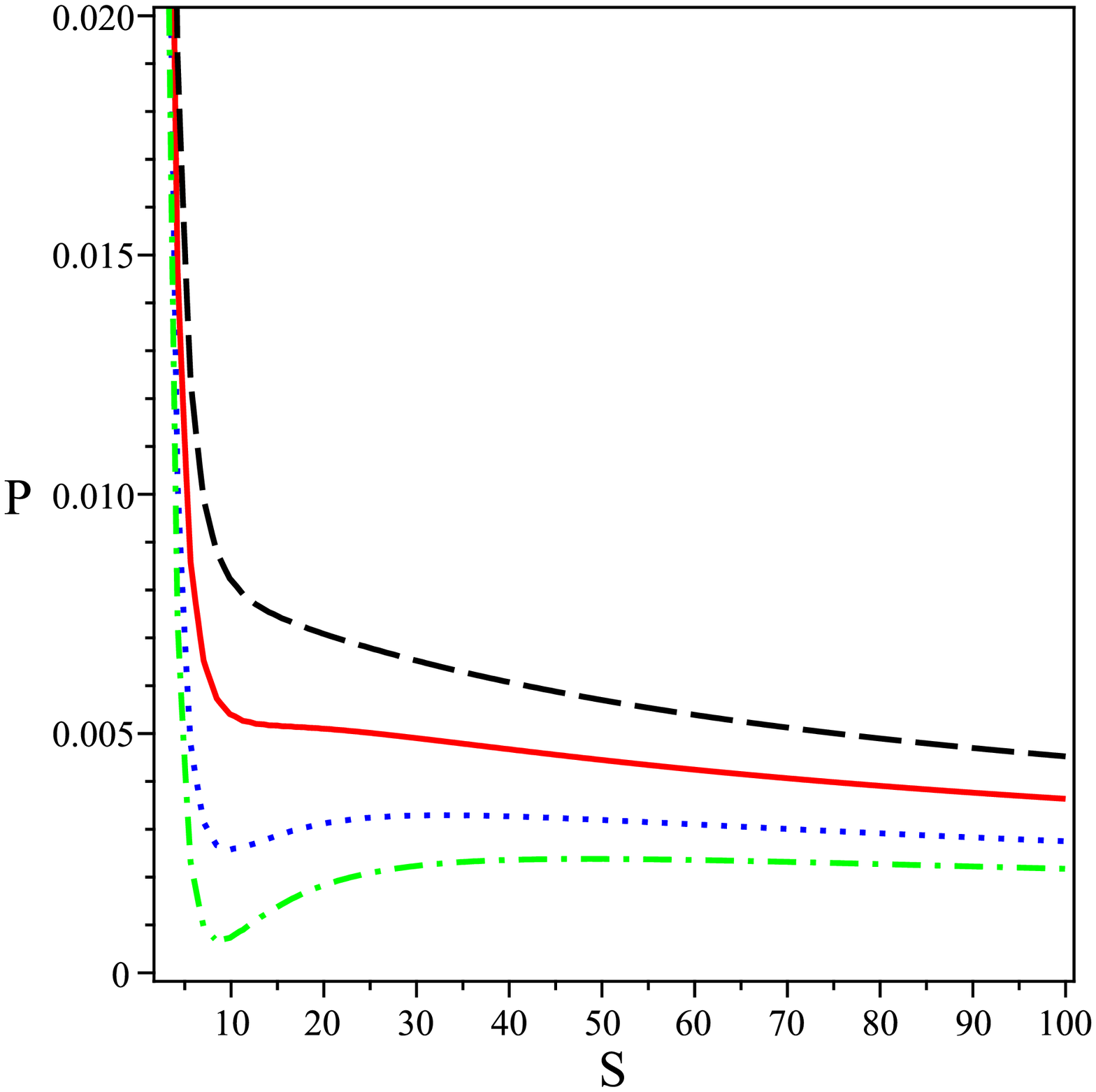}
 \label{figpr1}
 }
\subfigure[]  {\  \includegraphics[width=0.3\columnwidth]{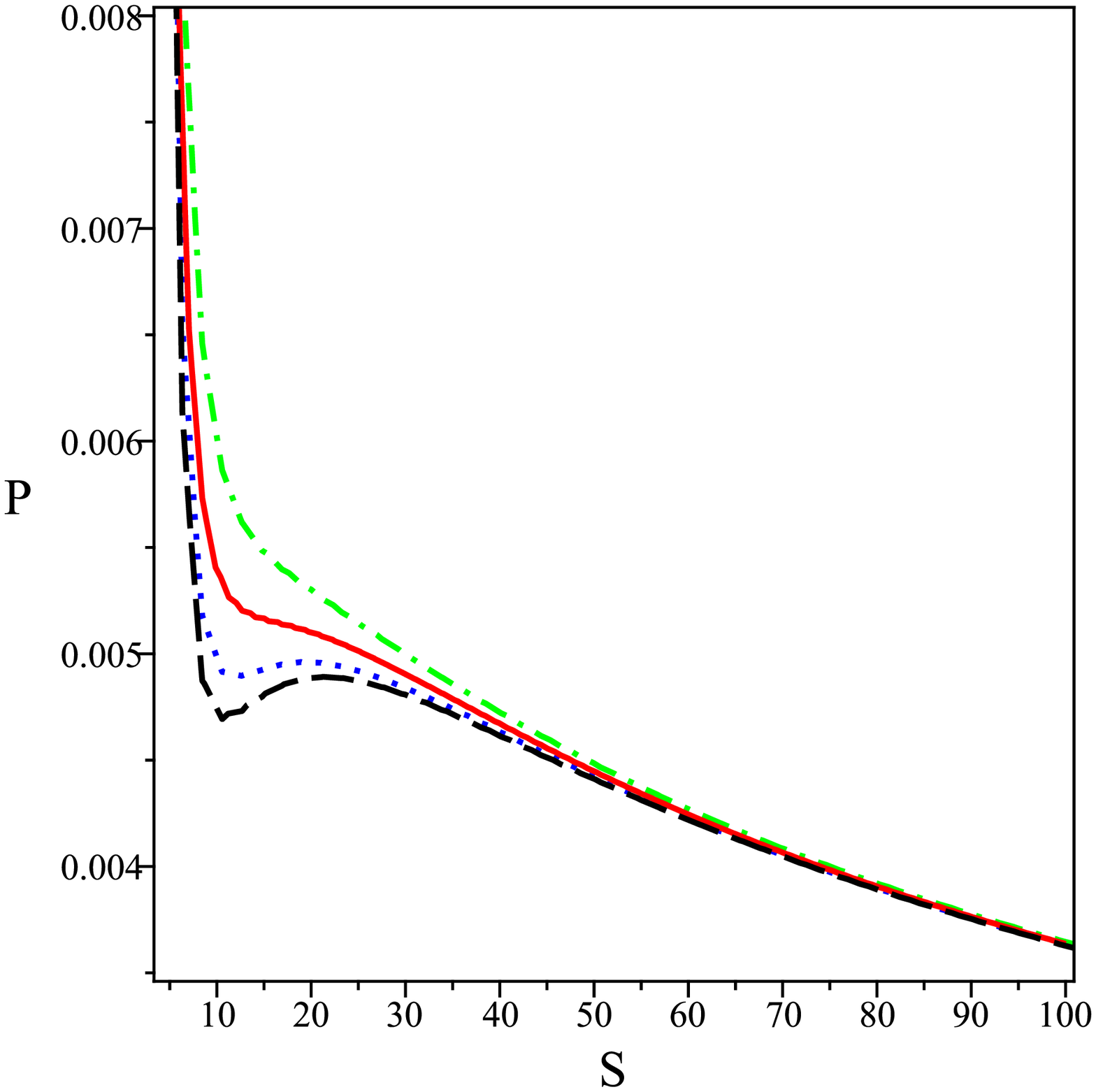}
\label{figpr2}  }
\subfigure[]  {\
\includegraphics[width=0.3\columnwidth]{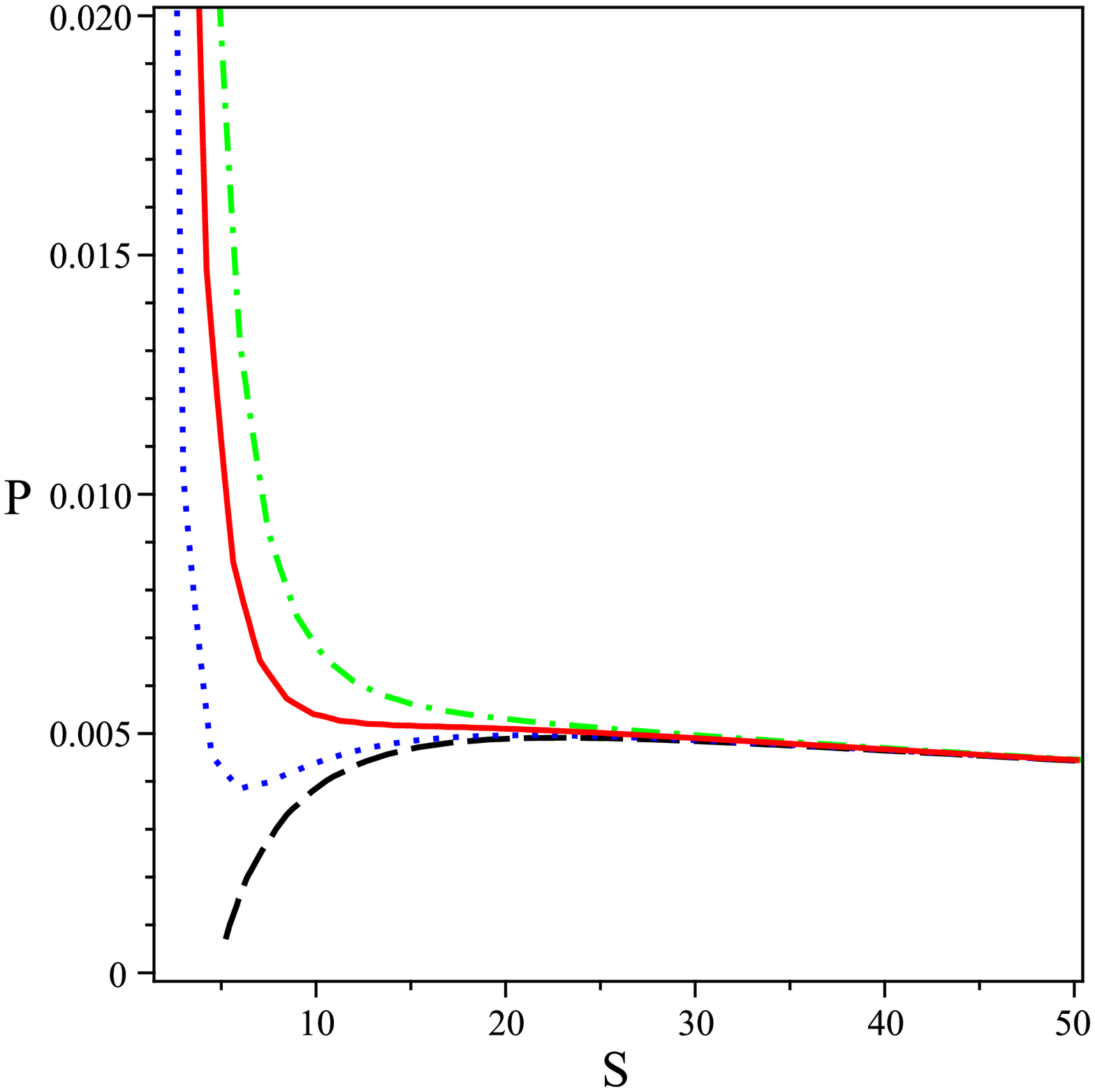}  \label{figpr3}  }
\caption{The behavior of $P$ in terms of $S $ for $T=0.038<T_{crit} $, $T=
\textcolor{blue}{0.045}<T_{crit}$, $T=T_{crit}=\textcolor{red}{0.055}$, $T=
\textcolor{black}{0.065}>T_{crit}$ and $J=0.1, Q=0.1$ (left), for $T=0.055$,
$J=0.5$ and $Q=\textcolor{black}{0.1},\textcolor{blue}{0.3},
\textcolor{black}{0.5},\textcolor{green}{0.7}$ (middle), for $T=0.055$, $%
Q=0.5$ and $J=\textcolor{black}{0.1},\textcolor{blue}{0.3},
\textcolor{red}{0.5},\textcolor{green}{0.7}$ (right).}
\label{prplot}
\end{figure}

By using Eq. (\ref{temp2}), we have plotted the pressure as a function of
the event horizon radius in Fig. (\ref{prplot}), keeping $T$, $Q$ and $J$
fixed. The temperature of isotherm diagrams decreases from top to bottom.
The upper black solid line corresponds to the ideal gas phase, the critical
isotherm is denoted by the red solid line, lower blue solid line corresponds
to temperature smaller than the critical temperature and below the
temperature the pressure becomes negative. Again, this is similar to the
pressure-volume plot in van der Waals liquid/gas system.

The isotherm corresponding to $T=T_{crit}$, has an inflection point. The
corresponding pressure and entropy at that point are called critical
pressure and critical entropy, respectively. So, the critical point is
determined by
\begin{equation}
\dfrac{\partial P}{\partial S}=0,\hspace{0.5cm}\dfrac{\partial ^{2}P}{%
\partial S^{2}}=0.
\end{equation}%
Using numerical methods for $Q=0.5,J=0.5$, we have $P_{crit}=%
\textcolor{red}{0.004802}$ and $S_{crit}=\textcolor{red}{15.64}$. Plugging
these values into Eq. (\ref{eqtem}), one finds $T_{crit}=%
\textcolor{red}{0.055}$. By using of above consequence, one can achieve an
approximately charge-independent result for the near universal ratio $\frac{%
P_{crit}\;S_{crit}}{T_{crit}}=\textcolor{red}{1.366}$.

\section{QNMs\label{QNMs}}

It is possible to obtain an analytic relation for the real and imaginary
parts of QNM as functions of the black hole charges ($M,a,q$) and of the
gravitational perturbation field ($j,m$). To do that, we obtain the field
equation for the mass less scalar probe as follows%
\begin{equation}
\nabla _{\alpha }\nabla ^{\alpha }\Phi =0.  \label{eqmotion}
\end{equation}%
By using of two Killing vectors of the metric $\partial _{t}$ and $\partial
_{\phi }$, one can use the separation of variables of the scalar field as%
\begin{equation}
\Phi (t,r,\theta ,\phi )=e^{-iwt+im\phi }R(r)S(\theta ).  \label{eqphi1}
\end{equation}%
Inserting Eq. (\ref{eqphi1}) into (\ref{eqmotion}) leads to two differential
equations for the angular $S(\theta )$ and radial $R(r)$ terms of wave
functions%
\begin{equation}
\dfrac{d}{dr}\left( \Delta _{r}\dfrac{dR(r)}{dr}\right) +\left( \dfrac{\Xi
^{2}\left( w^{2}(r^{2}+a^{2})^{2}+m^{2}a^{2}+2maw(\Delta
_{r}-r^{2}-a^{2})\right) }{\Delta _{r}}-E\right) R=0,  \label{eqradial}
\end{equation}%
\begin{equation}
\dfrac{d^{2}S(\theta )}{d\theta ^{2}}+\cot (\theta )\dfrac{dS(\theta )}{%
d\theta }+\dfrac{\Xi ^{2}}{\Delta _{\theta }}\left( -w^{2}a^{2}\sin
^{2}(\theta )-\dfrac{m^{2}}{\sin ^{2}(\theta )}+E\right) S(\theta )=0,
\label{eqangular}
\end{equation}%
where $E$ is a separation constant which can be obtained by using the
boundary conditions of regularity at $\theta =0$ and $\pi $. Since $E$ is a
function of $aw$, we can calculate it, perturbatively, for small $aw$ \cite%
{Casals:2018eev}
\begin{equation}
E=l(l+1)+2a^{2}w^{2}\left( \dfrac{m^{2}+l(l+1)-1}{(2l-1)(2l+3)}\right) +%
\mathcal{O}(a^{4}w^{4}),  \label{E}
\end{equation}%
where in the case of $aw\ll 1$, we can write $E=l(l+1)$ \cite{Casals:2018eev}%
. In order to study QNMs, first one can change the radial coordinate $r\in
(r_{+},\infty )$ to tortoise coordinate $x\in (-\infty ,\infty )$ as follows

\begin{equation}
\dfrac{dx}{dr}=\dfrac{r^{2}+a^{2}}{\Delta _{r}},\hspace{0.5cm}\psi (x)=\sqrt{%
r^{2}+a^{2}}R(r).  \label{tortoise}
\end{equation}%
Now, we can find that black hole perturbations from equation (\ref{eqradial}%
) can be reduced to the following second-order ODE%
\begin{equation}
\dfrac{d^{2}\psi (x)}{dx^{2}}+Q(x,w)\psi (x)=0,  \label{eqshrod}
\end{equation}%
where
\begin{equation}
Q(x,w)=\omega ^{2}+\dfrac{a^{2}m^{2}-4am\omega f(r)-E\Delta _{r}(r)}{%
(r^{2}+a^{2})^{2}}-\dfrac{r\Delta _{r}\Delta _{r}^{^{\prime }}}{%
(r^{2}+a^{2})^{3}}+\dfrac{\Delta _{r}^{2}(2r^{2}-a^{2})}{(r^{2}+a^{2})^{4}}.
\label{eqeff}
\end{equation}%
So, for the slowly rotating and small charged black hole $a\ll 1$, $\Lambda
=0$ and for the case $j\gg 1$ and $m\gg 1$, the effective potential becomes
\begin{equation}
Q=w^{2}-V=w^{2}-\left( 1-\dfrac{2M}{r}e^{-\dfrac{q^{2}}{2Mr}}\right) \dfrac{%
j(j+1)}{r^{2}}-\dfrac{4mMwa}{r^{3}}e^{-\dfrac{q^{2}}{2Mr}}.  \label{Q}
\end{equation}%
By using of Mashhoon's method \cite{vf,Mashhoon:1985cya} for QNMs, we can
obtain an approximate expression for quasinormal frequencies.

At first, we should note that the maximum of $V$ occurs at%
\begin{equation}
r_{ps}=3M\left( 1-\dfrac{2maw}{j(j+1)}\right) -\dfrac{2q^{2}}{3M}-\dfrac{%
17q^{4}}{216M^{3}}\left( 1+\dfrac{2maw}{j(j+1)}\right) ,  \label{eqrps}
\end{equation}%
and therefore, $r_{ps}$ is the radius of the photon sphere. The value of the
maximum potential is%
\begin{equation}
V_{m}=\dfrac{j(j+1)}{27M^{2}}\left( 1+\dfrac{q^{2}}{3M^{2}}+\dfrac{13q^{4}}{%
81M^{4}}\right) +\dfrac{4maw}{27M^{2}}\left( 1+\dfrac{q^{2}}{2M^{2}}+\dfrac{%
13q^{4}}{54M^{4}}\right)  \label{Vm}
\end{equation}%
and the curvature parameter is given by \cite{vf,Mashhoon:1985cya}%
\begin{equation}
\alpha =\dfrac{\sqrt{3}}{9M}\left( 1+\dfrac{5q^{2}}{18M^{2}}+\dfrac{23q^{4}}{%
216M^{4}}\right) +\dfrac{\sqrt{3}}{3M}\dfrac{2mwa}{j(j+1)}\left( 1+\dfrac{%
26q^{2}}{54M^{2}}+\dfrac{41q^{4}}{162M^{4}}\right) .
\end{equation}%
It should be noted that $w$ has a crucial role for all the above results. In
order to obtain $w$, we have inserted the above results in the equation (26)
of Ref. \cite{Mashhoon:1985cya}. Solving such a relation for $w$, one can
find the proper quasinormal frequencies as follows%
\begin{equation}
w=w_{r}+iw_{i}
\end{equation}%
where the real and imaginary parts of $w$ are, respectively,%
\begin{equation}
w_{r}=\dfrac{(j+\frac{1}{2})}{3\sqrt{3}M}\left( 1+\dfrac{q^{2}}{6M^{2}}+%
\dfrac{43q^{4}}{648M^{4}}\right) +\dfrac{2ma}{27M^{2}}\left( 1+\dfrac{q^{2}}{%
2M^{2}}+\dfrac{13q^{4}}{54M^{4}}\right) ,  \label{eqwr}
\end{equation}%
and%
\begin{equation}
w_{i}=\dfrac{\left( n+\frac{1}{2}\right) }{3\sqrt{3}M}\left( 1+\dfrac{5q^{2}%
}{18M^{2}}+\dfrac{23q^{4}}{216M^{4}}\right) +\dfrac{8ma}{27M^{2}j}\left( n+%
\dfrac{1}{2}\right) \left( 1+\dfrac{23q^{2}}{36M^{2}}+\dfrac{983q^{4}}{%
2592M^{4}}\right) .  \label{eqimagen}
\end{equation}%
It is noticable that all of the above results have been obtained in the case
of $j\gg 1,a\ll 1,q\ll 1$.

As we know, in the eikonal limit there is a relation between the quasinormal
frequencies and the properties of photon orbits. The real part of the
quasinormal frequencies is related to the angular velocity of the photon
orbit while the imaginary part is related to the Lyapunov exponent. In what
follows, we want to investigate the relation between the real part of
quasinormal modes and the radius of the shadow cast by the photon sphere of
the black hole. To do so, we use the Hamilton-Jacobi Method to obtain the
following equations of motion for null geodesics of charged rotating regular
black hole at equatorial plane (see Appendix \ref{app1}) \cite{takahashi,mnv}%
\begin{equation}
\Sigma \left( \dfrac{dr}{d\gamma }\right) =\sqrt{R(r)},\hspace{0.5cm}\Theta
(\theta )=\sqrt{k},\hspace{0.5cm}\Sigma \left( \dfrac{dt}{d\gamma }\right) ={%
T},\hspace{0.5cm}\Sigma \left( \dfrac{d\phi }{d\gamma }\right) =\varphi ,
\label{eqgeodesic}
\end{equation}%
where
\begin{equation}
R(r)=E^{2}\left( r^{2}+a^{2}-a\xi \right) ^{2}\left\{ 1-\frac{\left[
r^{2}+a^{2}-2f(r)\right] \left[ \eta +\left( \xi -a\right) ^{2}\right] }{%
\left( r^{2}+a^{2}-a\xi \right) ^{2}}\right\} ,  \label{R}
\end{equation}%
\begin{eqnarray}
T &=&E^{2}\left[ \dfrac{r^{2}+a^{2}}{\Delta }\left( r^{2}+a^{2}-a\xi \right)
-a\left( a-\xi \right) \right] , \\
\varphi &=&E^{2}\left[ \dfrac{a}{\Delta }\left( r^{2}+a^{2}-a\xi \right)
-\left( a-\xi \right) \right] ,
\end{eqnarray}%
and
\begin{equation}
\xi =\dfrac{L}{E},\hspace{0.5cm}\eta =\dfrac{k}{E^{2}}
\end{equation}%
are two impact parameters. The conditions for the unstable circular orbits
is given by $R(r)=dR(r)/dr=0$. Now, one can easily obtain the expressions
for $\xi $ and $\eta $ from the conditions of unstable circular orbits. For
the generic $f(r)$, these parameters take the following simple forms%
\begin{equation}
\xi =\dfrac{(a^{2}+r^{2})\left[ r+f^{^{\prime }}(r)\right] -4rf(r)}{a\left[
f^{^{\prime }}(r)-r\right] },  \label{eqxi}
\end{equation}%
\begin{equation}
\eta =\dfrac{r^{2}\left[ 8f(r^{2}+a^{2})+8rff^{^{\prime
}}-16f^{2}-2r^{3}f^{^{\prime }}-r^{2}f^{^{\prime }2}-4ra^{2}f^{^{\prime
}}-r^{4}\right] }{a^{2}(r-f^{^{\prime }})^{2}},  \label{eqeta}
\end{equation}%
where $r$ is the radius of the unstable circular orbits. These two equations
determine the contour of the shadow in $\xi -\eta $ plane. Furthermore, the
radius of the shadow is calculated as%
\begin{align}
R_{s}^{2}=\xi ^{2}+\eta = \dfrac{\left( 1+\frac{\left( a^{2}+2r^{2}\right)
q^{2}}{2\left( a^{2}-6r^{2}\right) Mr}\right) \left( a^{2}-6r^{2}\right)
M^{2}+a^{2}\left( q^{2}+2Mr\right) e^{\frac{q^{2}}{2Mr}%
}+r^{2}(a^{2}+2r^{2})e^{\frac{q^{2}}{Mr}}}{\left( M-re^{\frac{q^{2}}{2Mr}%
}\right) ^{2}}
\end{align}%
in which for the limit of $q\ll 1,a\ll 1$ and $j\gg 1$, we have%
\begin{equation}
R_{s}^{2}\approx \dfrac{2r^{2}(r^{2}-3M^{2})}{(r-M)^{2}}+\dfrac{4Mrq^{2}}{%
(r-M)^{2}}+\dfrac{(r^{2}+Mr-4M^{2})q^{4}}{2M(r-M)^{3}}.  \label{eqrs}
\end{equation}%
By inserting $w=\sqrt{w_{r}^{2}+w_{i}^{2}}$ into the Eq. (\ref{eqrps}), one
can obtain the radius of photon sphere as follows%
\begin{equation}
r=r_{ps}\approx 3M\left( 1-\dfrac{2q^{2}}{9M^{2}}-\dfrac{17q^{4}}{648M^{4}}%
\right) -\dfrac{2ma}{3\sqrt{3}jM}\left( 1+\dfrac{q^{2}}{6M^{2}}+\dfrac{%
17q^{4}}{648M^{4}}\right) .  \label{eqrps1}
\end{equation}%
It is obvious that in the limit $q\ll 1,a\ll 1$ and by inserting Eq. (\ref%
{eqrps1}) for $r_{ps}$ (at $j=m$) into the Eq. (\ref{eqrs}), we achieve%
\begin{equation}
{R_{s}=3\sqrt{3}M-4a-\dfrac{\sqrt{3}q^{2}}{2M}-\dfrac{q^{4}}{24\sqrt{3}M^{3}}%
+\dfrac{4aq^{4}}{3M^{4}},}
\end{equation}%
and it is easy to show that the real part of QNMs is inversely proportional
to the shadow radius as follows
\begin{equation}
w_{r}\approx \dfrac{j}{R_{s}}=\dfrac{j}{3\sqrt{3}M}\left( 1+\dfrac{q^{2}}{%
3M^{2}}+\dfrac{13q^{4}}{108M^{4}}\right) +\dfrac{4ma}{27M^{2}}\left( 1+%
\dfrac{2q^{2}}{3M^{2}}+\dfrac{35q^{4}}{108M^{4}}\right) ,\text{ for \ }j\gg 1
\label{eqshwr}
\end{equation}

\begin{center}
\begin{figure}[]
\hspace{4cm}\includegraphics[width=8.cm]{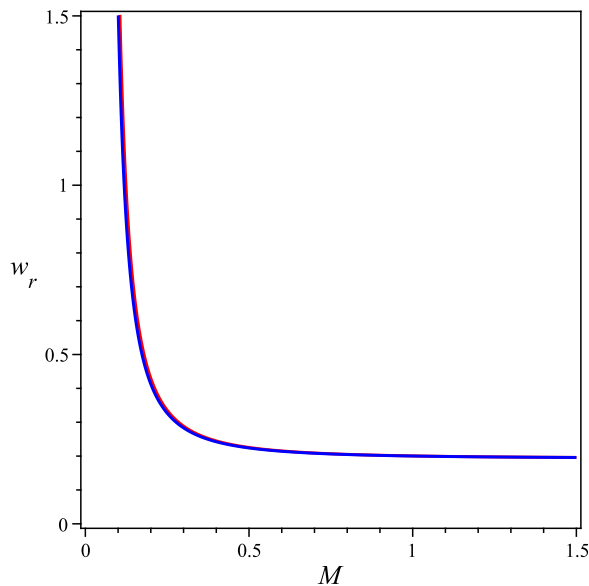}\vspace{0.1cm}
\caption{{\protect\small The comparison between $w_{r}$ in equation (\protect
\ref{eqshwr}) (red solid line) and equation (\protect\ref{eqwr}) (blue solid
line).}}
\label{gtplot}
\end{figure}
\end{center}

\begin{center}
\begin{figure}[]
\hspace{4cm}\includegraphics[width=8.cm]{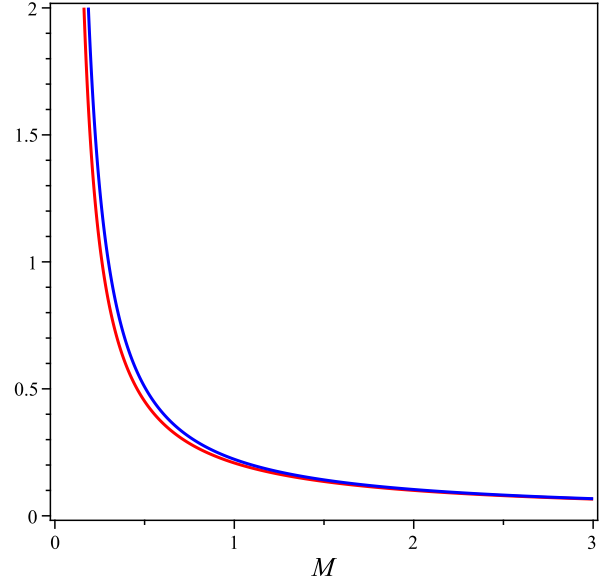}\vspace{0.1cm}
\caption{{\protect\small The comparison of $\protect\varsigma$ (red line)
and $w_i$ (blue line) in terms of $M $ for $a=q=0.1$.}}
\label{lambdaplot}
\end{figure}
\end{center}

In the following, we want to obtain the connection between Lyapunov exponent
and the imagenary parts of quasinormal mode (Eq. (\ref{eqimagen})). The
Lyapunov behavior exists in the systems that sensitive to the initial
conditions. So, the Lyapunov exponent which determines the value of the
sensitvity should have behavior looks like the imagenary parts of
quasinormal mode which determines instability of black hole. To extract the
Lyapunov exponent, we follow the method of Ref. \cite{Mashhoon:1985cya}. We
perturb the equatorial geodesic equations (\ref{eqgeodesic}) as%
\begin{equation}
r=r_{ps}\left[ 1+\epsilon f(t)\right] ,\hspace{0.5cm}\gamma =t+\epsilon h(t),%
\hspace{0.5cm}\phi =|w_{+}|\left[ t+\epsilon g(t)\right] .
\end{equation}%
Here, by perturbing the radial equation to leading order in $\epsilon $, we
obtain%
\begin{equation}
2\dot{t}^{2}\left. \dfrac{d^{2}f(t)}{dt^{2}}+\dfrac{d^{2}V}{dr^{2}}%
f(t)\right\vert _{r_{ps}}=0.  \label{eqft}
\end{equation}%
Now, by solving Eq. (\ref{eqft}) and using the condition $f(0)=0$, we can
find%
\begin{equation}
f(t)=\sinh (\varsigma t),\hspace{0.5cm}\left. \varsigma =\sqrt{-\dfrac{1}{{2%
\dot{t}^{2}}}{\frac{d^{2}V}{dr^{2}}}}\right\vert _{r_{ps}}.
\end{equation}%
So, by using of first equation for the radial and third equation for $t$ of
Eqs. (\ref{eqgeodesic}) and (\ref{eqxi}) and also the equation (\ref{eqrps})
for photon sphere radius in the limit of $q\ll 1,a\ll 1$ and $j\gg 1$, we
can write%
\begin{equation}
\varsigma =\dfrac{j}{3\sqrt{3}M}\left( 1+\dfrac{4q^{2}}{9M^{2}}+\dfrac{%
49q^{4}}{324M^{4}}\right) +\dfrac{8ma}{27jM^{2}}\left( 1+\dfrac{2q^{2}}{%
3M^{2}}+\dfrac{20q^{4}}{81M^{4}}\right) .
\end{equation}%
For the sake of completeness, we have done a comparison between Lyapunov
exponent and imaginary part of quasinormal frequency in Fig. \ref{lambdaplot}%
, and obviously, there is a good agreement between them.

\section{Conclusion\label{Con}}

In this study, we have addressed a rotating regular black hole in the
framework of Einstein's general relativity coupled to a nonlinear
electrodynamics. We have shown that for some values of parameters, there is
an event horizon for nonsingular solution and we can interpret it as a black
hole. In fact, we have obtained an upper limit for the magnetic charge or
rotation parameter in order to have event horizon and ergosphere. We have
plotted the ergoregion and have shown that as the magnetic charge and
rotation parameter increased, the ergoregion also increased. Since
thermodynamical behavior is of great importance in the search for a quantum
theory of gravitation, we have also managed to perform a thermodynamic
investigation of the black hole solutions. The conserved and thermodynamic
quantities have been calculated and the validity of the first law has been
examined. In addition, we have investigated the global stability of the
black hole by plotting the Gibbs free energy. Also, the heat capacity has
been studied to check the local stability. We have shown that the present
solutions admitted small/large phase transitions similar to the van der
Waals liquid/gas phase transition. Then, we have analytically studied the
quasinormal modes of black hole by using of Mashhoon's method. Finally, in
order to obtain a relation between the quasinormal frequencies and the
properties of the photon sphere, we have obtained the shadow radius of black
hole and Lyapunov exponent. As it can be seen, there is a good agreement
between the inverse of shadow radius and Lyapunov exponent with real and
imaginary parts of quasinormal modes, respectively.

\bigskip

\textbf{Acknowledgements}\newline
S. N. Sajadi acknowledge the support of Shiraz University.

\appendix

\section{Effective potential\label{app1}}

\begin{center}
\begin{figure}[]
\hspace{4cm}\includegraphics[width=8.cm]{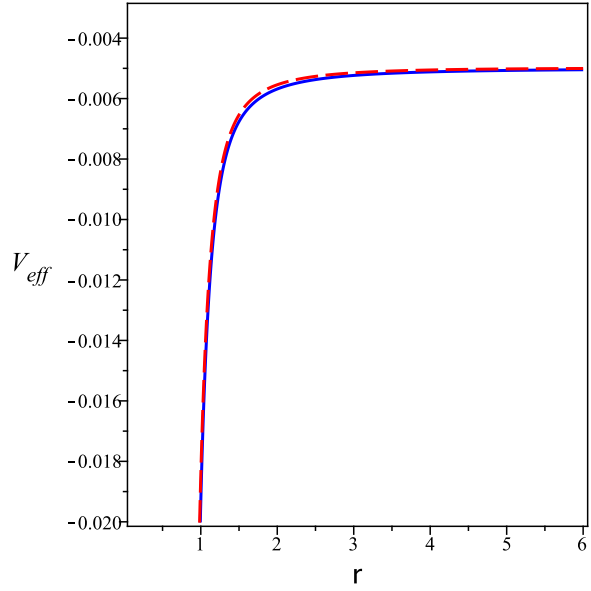}\vspace{0.1cm}
\caption{{\protect\small The behavior of effective potential in terms of $r $
for $M=1,\protect\theta=\frac{\protect\pi}{2},b=\frac{E}{L}=0.1,q=0.5,a=0.5 $%
. (\textcolor{red}{dashed line} for metric (\protect\ref{metric}) and
\textcolor{blue}{solid line} for metric (\protect\ref{eq22}).}}
\label{effec}
\end{figure}
\end{center}

Here, we want to show why we have used the original metric instead of
effective metric to study the shadow of black hole. As we know, photons in
the nonlinear electrodynamics do follow the null geodesics of an effective
metric rather than of the original one \cite{mnv}. So, we need to study the
effective metric. The effective geometry has been written as follows
\begin{equation}
g_{eff}^{\mu \nu }=L_{F}g^{\mu \nu }-4L_{FF}F_{\alpha }^{\mu }F^{\alpha \nu
}.
\end{equation}%
So, the effective metric in the plane $\theta =\dfrac{\pi }{2}$, can be
written as
\begin{equation}
ds^{2}=g_{tt}^{eff}dt^{2}+g_{rr}^{eff}dr^{2}+2g_{t\phi }^{eff}dtd\phi
+g_{\phi \phi }^{eff}d\phi ^{2},  \label{eq22}
\end{equation}%
where
\begin{equation}
g_{tt}^{eff}=\dfrac{-{L_{F}}{r}^{6}+2{L_{F}}f{r}^{4}+4{L_{FF}}{a}^{2}{q}%
^{2}+4{L_{FF}}{q}^{2}{r}^{2}-8{L_{FF}}f{q}^{2}}{{L_{F}}{r}^{2}\left( {L_{F}}{%
r}^{4}-4{L_{FF}}{q}^{2}\right) },
\end{equation}%
\begin{equation}
g_{rr}^{eff}={\dfrac{{r}^{2}}{{L_{F}}\left( {a}^{2}+{r}^{2}-2f\right) }},
\end{equation}%
\begin{equation}
g_{t\phi }^{eff}=-{\dfrac{2a\left( -{L_{F}}f{r}^{4}-2{L_{FF}}{a}^{2}{q}^{2}-2%
{L_{FF}}{q}^{2}{r}^{2}+4{L_{FF}}f{q}^{2}\right) }{{r}^{2}{\ L_{F}}\left( -{\
L_{F}}{r}^{4}+4{L_{FF}}{q}^{2}\right) }},
\end{equation}%
\begin{equation}
g_{\phi \phi }^{eff}={\dfrac{-{L_{F}}{a}^{2}{r}^{6}-{L_{F}}{r}^{8}-2{L_{F}}{a%
}^{2}f{r}^{4}-4{L_{FF}}{a}^{4}{q}^{2}-4{\ L_{FF}}{a}^{2}{q}^{2}{r}^{2}+8{%
L_{FF}}{a}^{2}f{q}^{2}}{{r}^{2}{L_{F}}\left( -{L_{F}}{r}^{4}+4{L_{FF}}{q}%
^{2}\right) }}.
\end{equation}%
Here $L_{F}$ is the same as that of Eq. (\ref{eqlaf}) but at $\theta =\dfrac{%
\pi }{2}$ and $L_{FF}=\dfrac{dL_{F}}{dF}$ at $\theta =\dfrac{\pi }{2}$. It
is useful to study the effective potential that is felt by the photons. By
using of the symmetries of the metric one can achieve to the following
equation
\begin{equation}
\dfrac{1}{2}\dot{r}^{2}+V_{eff}=\Xi
\end{equation}%
where
\begin{equation}
V_{eff}=\dfrac{1}{2g_{rr}}\left( g_{tt}\dot{t}^{2}+2g_{t\phi }\dot{t}\dot{%
\phi}+g_{\phi \phi }\dot{\phi}^{2}\right) ,\hspace{0.5cm}E=g_{\alpha \beta
}\xi _{t}^{\alpha }\dot{x}^{\beta },\hspace{0.5cm}L=g_{\alpha \beta }\xi
_{\phi }^{\alpha }\dot{x}^{\beta }
\end{equation}%
and $g_{ab}$ are the components of the metric, $\xi $ is the Killing vector
and $\Xi $ is the total energy. Considering Fig. \ref{effec}, we give plots
of $V_{eff}$ for the metric (\ref{eq22})(blue line) and for metric (\ref%
{metric})(red line). One can see that the difference between the two plots
is insignificant and it is thus legitimate to use the original metric
instead of effective one.

\end{document}